\documentclass[%
preprint,
 amsmath,amssymb,
 aps,
]{revtex4-2}

\usepackage{graphicx}% Include figure files
\usepackage{dcolumn}% Align table columns on decimal point
\usepackage{bm}% bold math
\usepackage{color}

\begin{document}

%\preprint{APS/123-QED}

\title{The effect of demographic stochasticity on predatory-prey oscillations}% Force line breaks with \\
%\thanks{A footnote to the article title}%

\author{Solmaz Golmohammadi}
\affiliation{%
Department of Physics, Institute for Advanced Studies in Basic Sciences (IASBS), Zanjan, 45137-66731, Iran
}%
\affiliation{
The Abdus Salam International Centre for Theoretical Physics (ICTP), Strada Costiera 11, 34014 Trieste, Italy
}%

\author{Mina Zarei}%
 \email{mina.zarei@iasbs.ac.ir}
\affiliation{%
Department of Physics, Institute for Advanced Studies in Basic Sciences (IASBS), Zanjan, 45137-66731, Iran
}%

\author{Jacopo Grilli}
 \email{jgrilli@ictp.it}
\affiliation{
Quantitative Life Sciences section, The Abdus Salam International Centre for Theoretical Physics (ICTP), Strada Costiera 11, 34014 Trieste, Italy
}%

%\date{\today}% It is always \today, today,
             %  but any date may be explicitly specified

\begin{abstract}
The ecological dynamics of interacting predator and prey populations can display sustained oscillations, as for instance predicted by the Rosenzweig-MacArthur predator-prey model. The presence of demographic stochasticity, due to the finiteness of population sizes, alters the amplitude and frequency of these oscillations. 
Here we present a method for characterizing the effects of demographic stochasticity on the limit cycle attractor of the Rosenzweig-MacArthur. We show that an angular Brownian motion well describes the frequency oscillations. In the vicinity of the bifurcation point, we obtain an analytical approximation for the angular diffusion constant. 
This approximation accurately captures the effect of demographic stochasticity across parameter values.
%We compared the results obtained by the stochastic simulation of the predator-prey model with those of the analytical method by  varying the input parameters. The good agreement in the results does illustrate that the proposed method is accurate to quantify the demographic stochasticity along the limit cycle.
%\begin{description}
%\item[Usage]
%Secondary publications and information retrieval %purposes.
%\item[Structure]
%You may use the \texttt{description} environment %to structure your abstract;
%use the optional argument of the \verb+\item+ %command to give the category of each item. 
%\end{description}
\end{abstract}

%\keywords{Suggested keywords}%Use showkeys class option if keyword
                              %display desired
\maketitle

%\tableofcontents

%\jg{TODO: \\
%add references 
%}

\section{\label{sec:level1}Introduction}

The independent work of Alfred Lotka and Vito Volterra~\cite{lotka1920analytical, volterra1927variazioni} demonstrated over a century ago how predator-prey population dynamics can exhibit neutral oscillation, where the amplitude depends on the initial condition, with a fixed frequency. The critical assumption in the original formulation, ultimately giving rise to neutral oscillations, is that the growth rate of predators depends linearly on the density of preys (so-called linear functional responses). Their models were extended by Rosenzweig and MacArthur~\cite{rosenzweig1963graphical} to incorporate more realistic saturating functional responses (Holling type II~\cite{holling1959components}). In this setting, the dynamics converge to a stable limit cycle in certain parameter regimes. In that case, prey and predator populations oscillate with fixed frequency and amplitude, independently of initial conditions. Population oscillations have been observed in natural predator-prey ecosystems and confirmed by experiments~\cite{gilpin1973hares, elton1942ten, butler1953nature, korpimaki2005predator, higgins1997stochastic, kamata2000population}, although there is ongoing debate regarding whether their origins are endogenous, as predicted by the Rosenzweig-MacArthur (RMA) model, or exogenous, arising for instance from periodic (e.g., seasonal) environmental fluctuations.

In reality, population dynamics are generally stochastic due to intrinsic or extrinsic factors. Extrinsic factors include environmental fluctuations such as weather, fires, and seasonality, as well as biotic disturbances. Earlier studies have examined the stochastic behavior of population dynamics and the impact of external noise on these dynamics~\cite{may1973stability,melbourne2008extinction,Petrovskii2010,reuman2008colour}

Intrinsic factors arise from demographic effects, where the discrete nature of populations leads to random and independent birth, death, and migration processes. Accurately capturing this inherent stochasticity is essential for understanding the complex dynamics and interactions that govern ecological systems. Various methods and formalisms have been developed to address this stochasticity, each with its own strengths and limitations tailored to specific research questions. These approaches can be explored both analytically and numerically, providing a comprehensive view of the dynamics within these intricate systems~\cite{black2010stochastic}. 

In our study, we present a novel analytical framework that models the noise affecting the phase of the limit cycle in a stochastic predator-prey system as angular Brownian motion. By approximating the limit cycle as an ellipse through the Van Kampen expansion of the master equation, we derive key insights into the system's behavior. Our findings are further validated through simulations using the Gillespie algorithm applied to a stochastic Rosenzweig-MacArthur model. This dual methodology, which integrates robust numerical simulations with analytical techniques, underscores the advantages of Individual-Based Models (IBMs) over traditional population-level and agent-based models, offering a deeper understanding of the effects of demographic stochasticity on predator-prey dynamics.

In section~\ref{sec:II}, we introduce the stochastic Rosenzweig-MacArthur model. In Section~\ref{sec:III} we present our results. First, we give an analytical approximation to find the mean frequency and the radius of the limit cycle close to the bifurcation point in the determining case. Second, we show that the perturbations along the limit cycle generated by demographic stochasticity can be approximated by an angular Brownian motion. Finally, we extract an analytic expression for the noise strength using Van Kampen system size expansion of the corresponding Fokker-Planck equation. Section~\ref{sec:IV} presents discussion and concluding remarks. 

\section{\label{sec:II} The Rosenzweig-MacArthur Predator-Prey Model}

The Rosenzweig-MacArthur Predator-Prey (RMA) Model~\cite{rosenzweig1963graphical} provides a standard framework to study the coupled dynamics of predator and prey populations. The population abundances of the prey ($R$) and predators ($F$) change in time accordingly to
\begin{eqnarray}
&& \dot{R} = aR-\dfrac{R^2}{2N}- \dfrac{sRF}{1+ s \tau R}, \nonumber \\  
&& \dot{F} = -d F+ \dfrac{sRF}{1+ s \tau R} \ .
\label{eq001}
\end{eqnarray} 
In the absence of predators ($F=0$), the preys grow logistically, with growth rate $a$ and carrying capacity $2 N a$.
In the absence of prey ($R=0$) the predator population declines with the death rate $d$. The per-capita predation rate has a Holling type II functional response which saturates to $1/\tau$ for large predator populations. The parameter $\tau$ is the handling time of each prey. The base predation rate is equal to $s$.

%Population dynamics is one of the most widely studied topics within Bio-mathematics. Oscillation in the dynamics of animal populations has been observed in nature and demonstrated in laboratory~\cite{raynal2019modeles}. One of the famous equations presented for modeling these cyclic relationships is Rosenzweig-MacArthur (RMA) model. In the following, we will briefly review the general architecture of the deterministic and stochastic RMA models.

%\subsection{\label{sel. ec:deter}Deterministic RMA Model}
Without loss of generality, we define $d=1$ (which corresponds to set the timescale of the system to be measured relatively to the predators lifespan) and set population abundances $x=R/N$ and $y=F/N$ to obtain
\begin{eqnarray} 
%\begin{split}
&& \dot{x} = ax - \dfrac{x^2}{2}- \dfrac{\sigma xy}{1+ \sigma \tau x}, \nonumber \\ 
&& \dot{y} = -y + \dfrac{\sigma xy}{1+ \sigma \tau x} \ ,
%\end{split}
\label{eq002}
\end{eqnarray}
where $\sigma = sN$. 

\begin{figure}
\begin{minipage}{.49\hsize}
\includegraphics[width=\linewidth]{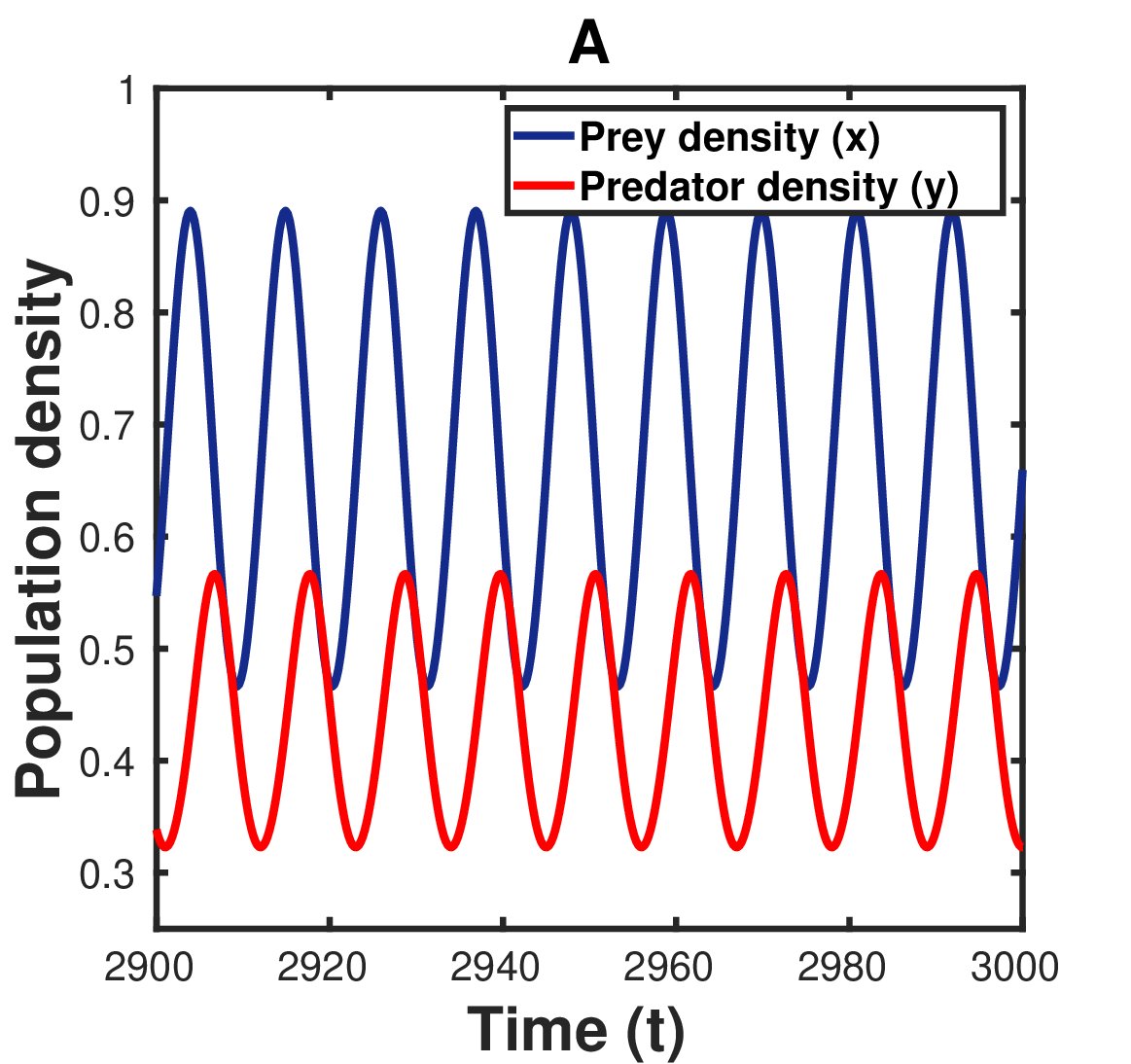}
\end{minipage}
\begin{minipage}{.49\hsize}
\includegraphics[width=\linewidth]{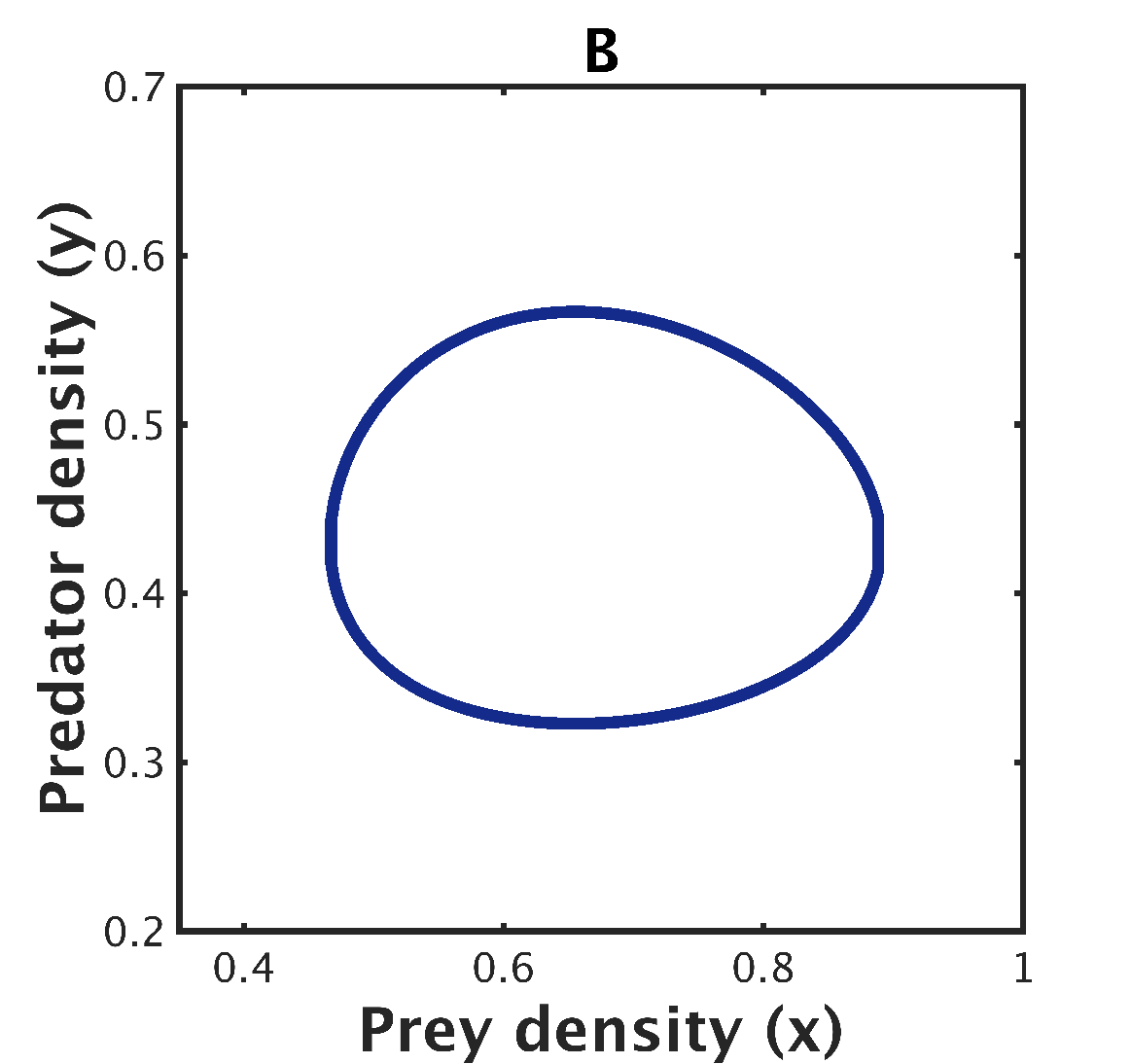}
\end{minipage}
\caption{Deterministic RMA model. (A) Oscillations of populations versus time, where blue and red correspond to prey and predator densities respectively. (B) The corresponding limit cycle  in phase space. Parameters in these figures are $\tau = 0.5$, $a = 1$ and $\sigma = 3.05$.
}
\label{fig:1}
\end{figure}

Equation \ref{eq002} admits three distinct fixed points~\cite{rosenzweig1963graphical, doi:10.1137/0512047, smith2016extinction}
\begin{eqnarray} 
%\begin{split}
&& M_{1} = \left(x=0, y=0\right), \nonumber \\
&& M_{2} = \left(x=2a, y=0\right),\nonumber \\
&& M_{3} = \left(x=\dfrac{1}{\sigma(1-\tau)}, y=\dfrac{2a\sigma(1-\tau)-1}{2\sigma^{2}(1-\tau)^{2}} \right) \ .
%\end{split}
\label{eq0021}
\end{eqnarray}
The trivial fixed point $M_1$ corresponds to the extinction of both species and is always unstable, as, in absence of predators, preys are always able to invade. $M_2$ corresponds to the state that only the prey is present. This point is also unstable for
where
\begin{eqnarray} 
%\begin{split}
\sigma > \sigma_{0}= \dfrac{1}{2a(1-\tau)} \ .
%\end{split}
\label{eq0021}
\end{eqnarray}
Under this condition, the predators are able to grow on a population of preys with density equal to its carrying capacity.
The third fixed point $M_3$ corresponds to coexistence between predators and preys and it is stable if
\begin{eqnarray} 
%\begin{split}
\sigma_{0}< \sigma < \sigma^{*} = \dfrac{1+ \tau}{2a\tau (1-\tau)} \ .
\label{eq0021b}
\end{eqnarray}
For $\sigma < \sigma_0 $ the system converge to $M_2$. For $\sigma > \sigma^{*}$, $M_3$ is an unstable (as all the other fixed points) and the dynamics converges to a stable limit cycle as depicted in figure~\ref{fig:1}.

We generalize this deterministic model to include stochasticity by considering a discrete number of individuals $F$ and $R$, which change with the following rates:
%\begin{eqnarray} 
%\begin{split}
%&& T(F,R+1|R,F) = a R, \nonumber \\
%&& T(F+1,R|R,F) = \eta \dfrac{sRF}{1+ s \tau R} ,\nonumber \\
%&& T(F,R-1|R,F) =  \dfrac{sRF}{1+ s \tau R} + %\dfrac{R^2}{2N} ,\nonumber \\
%&& T(F-1,R|R,F) = F,\nonumber \\
%&&  \ ,
%\end{split}
%\label{eq0_rates}
%\end{eqnarray}
\begin{eqnarray} 
%\begin{split}
&& T(F,R+1|R,F) = a R, \nonumber \\
&& T(F+1,R-1|R,F) =\dfrac{sRF}{1+ s \tau R} ,\nonumber \\
&& T(F,R-1|R,F) =   \dfrac{R^2}{2N} ,\nonumber \\
&& T(F-1,R|R,F) = F,
%\end{split}
\label{eq0_rates2}
\end{eqnarray}
where $T(C_f|C_i)$ indicates the rate from initial configuration $C_i$ to final configuration $C_f$.
%\jg{I added this. Check it is correct ($\eta$?) if $\eta$ is always $1$ we should change eq.~\ref{eq001} by removing $\eta$ and the new structure of the rates would be
%\begin{eqnarray} 
%\begin{split}
%&& T(F,R+1|R,F) = a R, \nonumber \\
%&& T(F+1,R-1|R,F) =\dfrac{sRF}{1+ s \tau R} ,\nonumber \\
%&& T(F,R-1|R,F) =   \dfrac{R^2}{2N} ,\nonumber \\
%&& T(F-1,R|R,F) = F,\nonumber \\
%&&  \ ,
%\end{split}
%\label{eq0_rates2}
%\end{eqnarray}
%}
%

%In fact, stochastic models are generally a better fit to the observations than the deterministic ones~\cite{gardiner1985handbook}.
%The Gillespie algorithm is a stochastic method enables sampling out of probability distributions described by master equations. 
These rates fully define the stochastic process that determines the dynamics of the system~\cite{smith2016extinction}. In the limit of $N \to \infty$, demographic stochasticity can be neglected and the trajectories can be approximated as the solutions of eq.~\ref{eq002}.
%Events in the stochastic RMA system and their rates are described in the table~\ref{tab1}.   

%\begin{table}[]

%\centering
%\caption{Events and their rates in the stochastic RMA model.}
%\begin{tabular}{|l|l|}
%\hline
%Event                      & Rate                    \\ \hline
%Birth of rabbit            & $aR$                    \\ \hline
%Computation among rabbits  & $\dfrac{R^2}{2N}$       \\ \hline
%Predation and birth of fox & $\dfrac{sRF}{1+s\tau R}$ \\ \hline
%Death of fox               & $-F$                    \\ \hline
%\end{tabular}
%%\label{tab1}
%\end{table}

\begin{figure}
\begin{minipage}{.75\hsize}
\includegraphics[width=\linewidth]{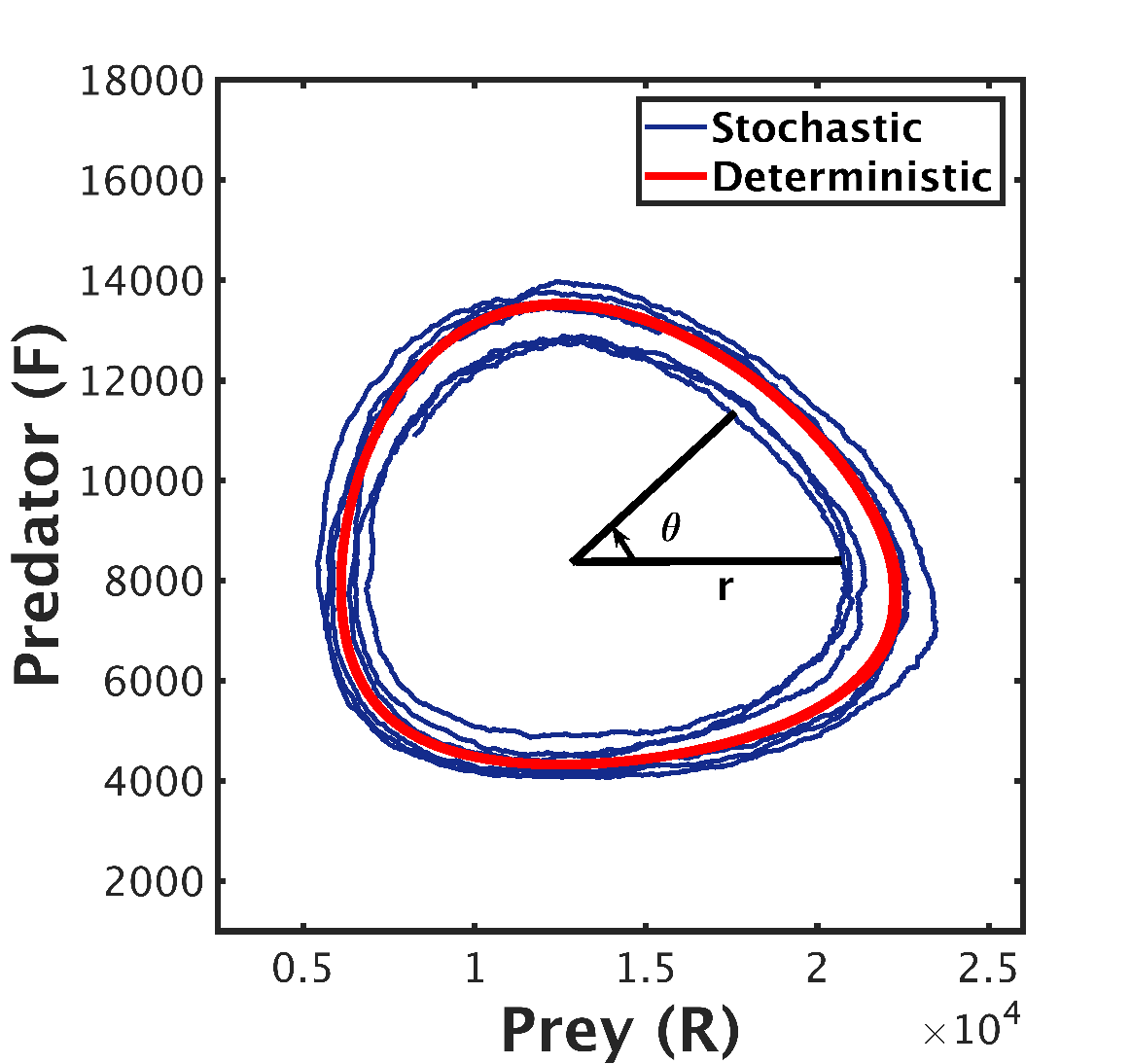}
\end{minipage}
%\begin{minipage}{.49\hsize}
%\includegraphics[width=\linewidth]{ONE-B.jpg}
%\end{minipage}
\caption{Limit cycles of the deterministic and Stochastic RMA models.  Radius ($r$) and phase ($\theta$) in the phase space has been determined in the figure. Deterministic limit cycle has been distinguished by red color. Parameters in these figures are $\tau = 0.5$, $a = 1$, $N=20000$ and $\sigma = 3.2$. }
\label{fig:2}
\end{figure}

Figure~\ref{fig:2} displays one stochastic trajectory of the RMA system (generated using the Gillespie algorithm) in the parameter regime where a limit cycle in the deterministic system is expected. While the trajectories are noisy, they still resemble the deterministic limit cycle.

\section{\label{sec:III}Results}

In the previous section, we provided the definition of the deterministic and stochastic RMA model. In the parameter regime where a limit cycle occurs, the stochastic trajectory still displays an oscillatory behavior.
In this section, we linearize the fluctuations around the deterministic attractor and quantify the effect of stochasticity. We then compare the analytical results to numerical simulations.  

\subsection{\label{sec:III-A}Elliptical Approximation of the Limit Cycle}

In order to quantify the noise strength along the limit cycle, it is convenient to rewrite the differential equations in the polar phase space. We can approximate the shape of the limit cycle to an ellipse, when the amplitude of the oscillations is small. In particular, this occurs when the parameter $\sigma / \sigma^* \to 1^+$, and in that case the oscillations occur in the vicinity of the $M_3$. It is convenient to introduce $\gamma = \dfrac{\sigma - \sigma^{*}}{\sigma^{*}}$. Our approximation holds for small and positive values of $\gamma$.
We define $dx$ and $dy$ as follows:
\begin{eqnarray}
&& dx = r(t)\cos\theta(t), \nonumber \\
&& dy = r(t)\epsilon \sin\theta(t),
\label{eq3}
\end{eqnarray}
where $dx=(x-x_{f})$, $dy=(y-y_{f})$ and $M_3 = (x_{f}, y_{f})$. The variable $\theta$ is the phase variable on the limit cycle centered at the mentioned fixed point~($M_3$), which is calculated as follows: 
\begin{equation}
\theta = \text{tan}^{-1}\left(\dfrac{y-y_{f}}{x-x_{f}}\right).
\label{eq003}
\end{equation}
The phase value goes from $0$ to $2\pi$ as the system evolves by time and completes a cycle in the phase-space, as illustrated in Figure~\ref{fig:2}.
By expanding the equation~\ref{eq002} around $(x_{f}, y_{f})$ and using equation~\ref{eq3}, one can find $\dot{r}$ and $\dot{\theta}$ (see details in the supplementary materials). Under this approximation, we obtained the following expressions for the mean frequency ($\omega$) and the mean radius ($\left\langle r \right\rangle$)
\begin{eqnarray}
&& \omega = \epsilon =  \sqrt{a(1-\tau)- \dfrac{1}{2\sigma}},  \nonumber \\
\nonumber \\
&& \left\langle r \right\rangle = \dfrac{2a\sqrt{\dfrac{\gamma \tau(1+\tau)}{1+\gamma-\tau+\gamma \tau}}}{1+\gamma+\tau+\gamma \tau}  \ .
\label{eq6}
\end{eqnarray}

\begin{figure}
\begin{minipage}{.49\hsize}
\includegraphics[width=\linewidth]{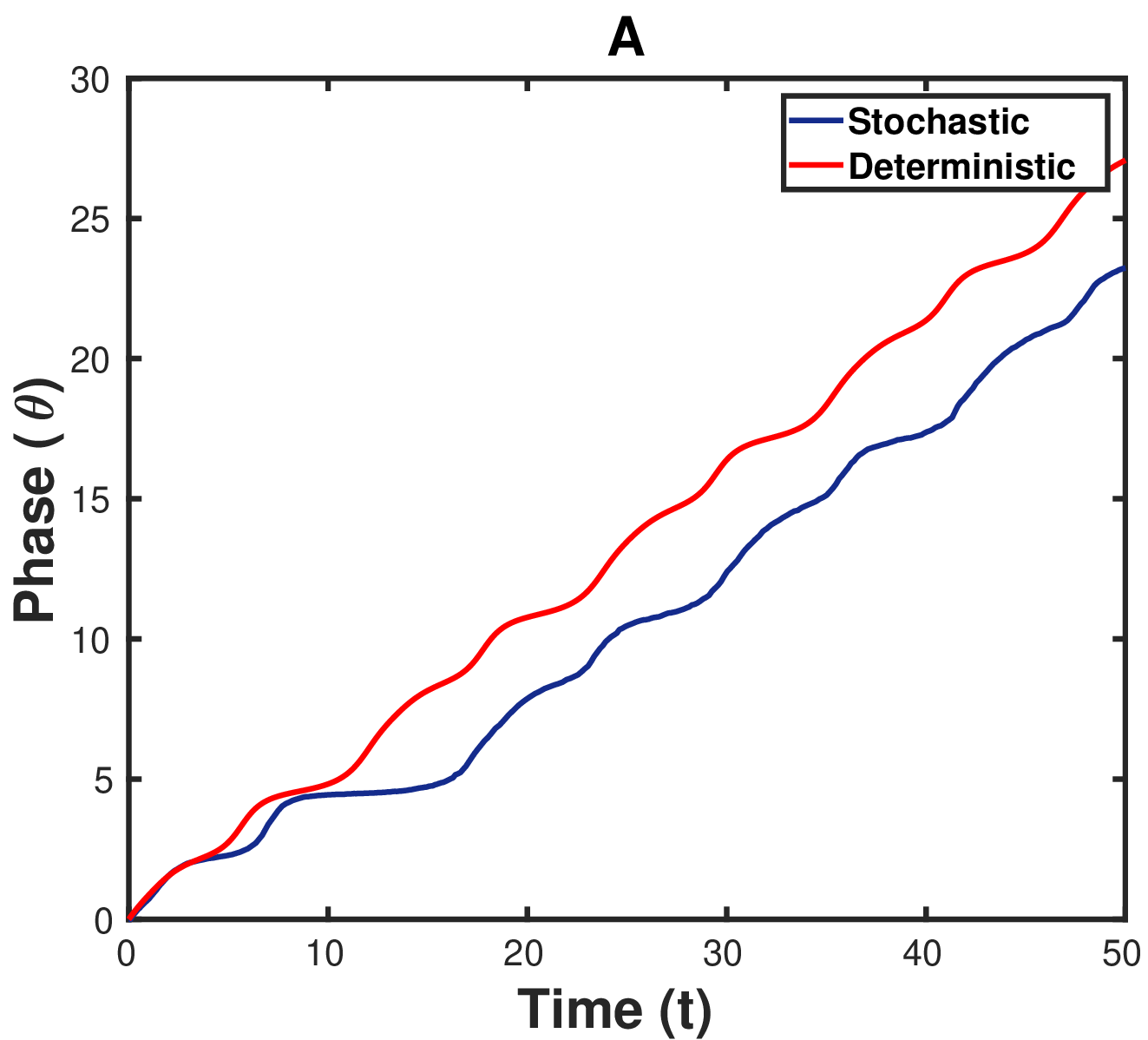}
\end{minipage}
\begin{minipage}{.49\hsize}
\includegraphics[width=\linewidth]{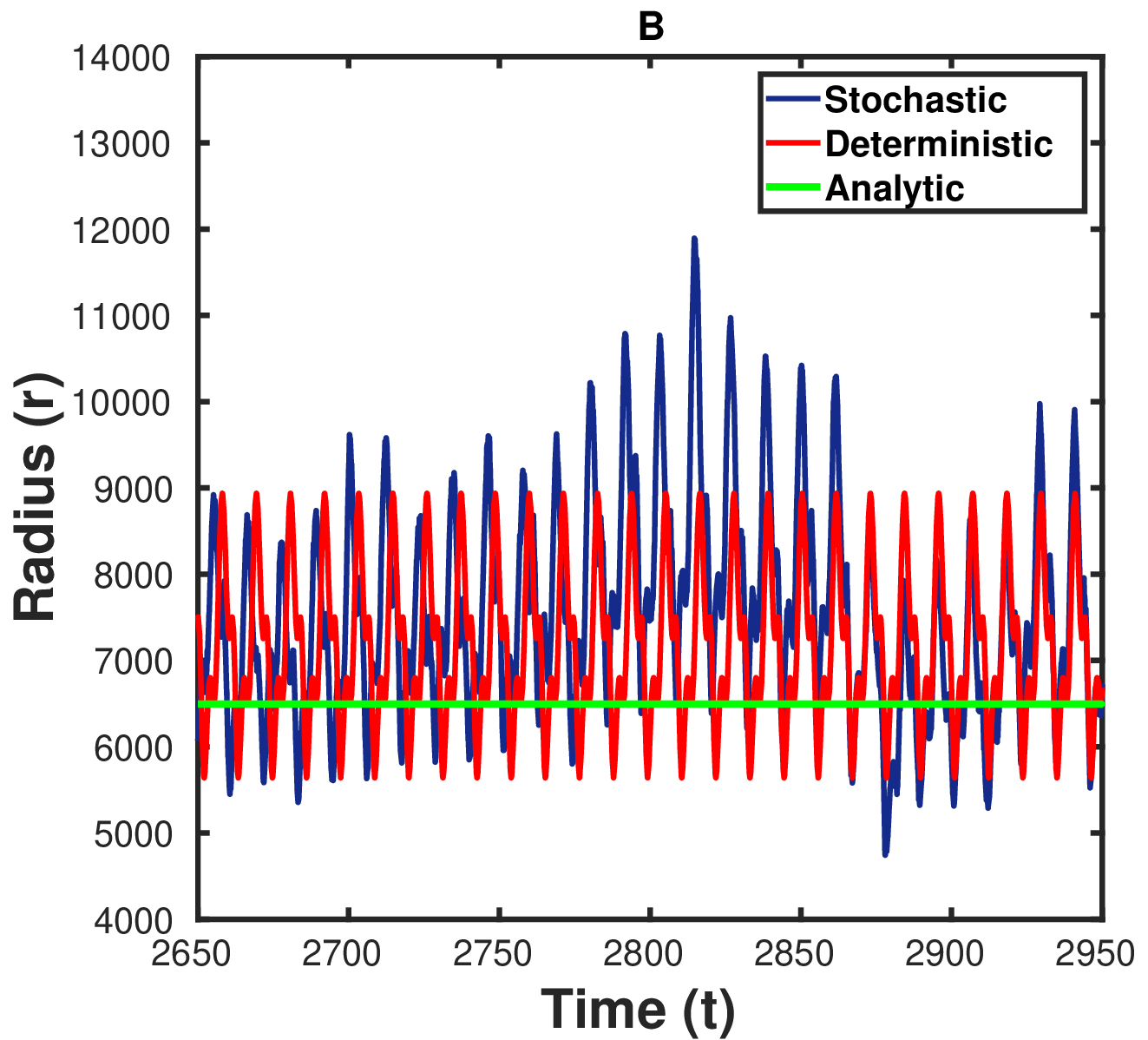}
\end{minipage}
\caption{Comparison of deterministic, stochastic and analytic approximation of RMA model. (A) Phase versus time (N = 1800). (B) Radius versus time (N = 20000). Parameters in these figures are $\tau = 0.5$, $a = 1$ and $\sigma = 3.2$.}
\label{fig:3}
\end{figure}

 We provide a confirmation of the analytical results using direct numerical simulation of the deterministic and of the stochastic RMA models. Figure~\ref{fig:3} compares the time evolution of the phase and radius of the limit cycle. Note that, due to the fact that the limit-cycle does not have a perfect elliptic shape, we observe oscillations of both radius and frequency of oscillations.
 Nevertheless, our analytical approximation for mean radius is in agreement with the simulation results. %\textbf{ We have not shown the agreement of analyical frequency. \jg{is it easy to add it?}}
 
\begin{figure}[h!]
\begin{minipage}{.49\hsize}
\includegraphics[width=\linewidth]{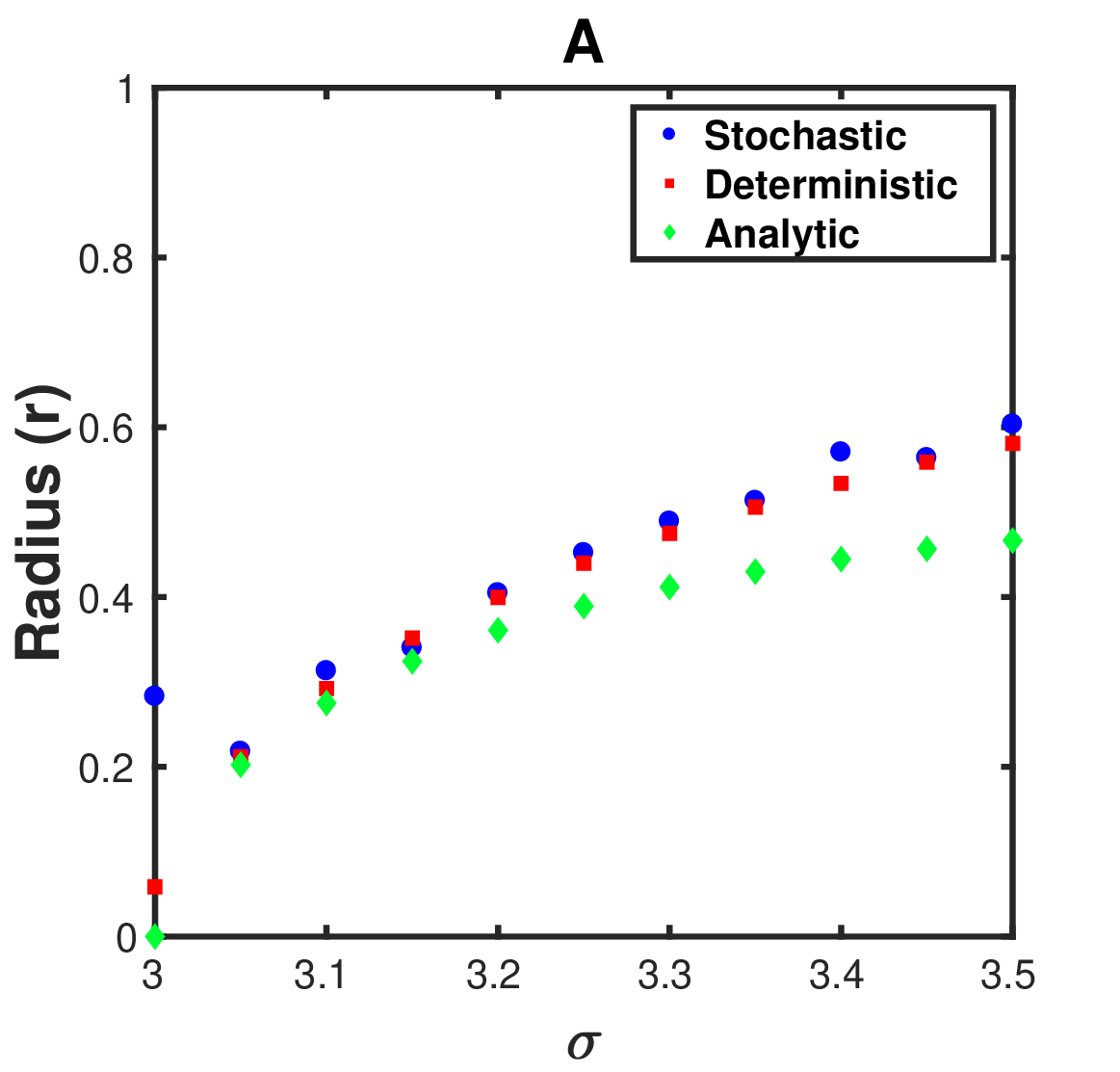}
\end{minipage}
\begin{minipage}{.49\hsize}
\includegraphics[width=\linewidth]{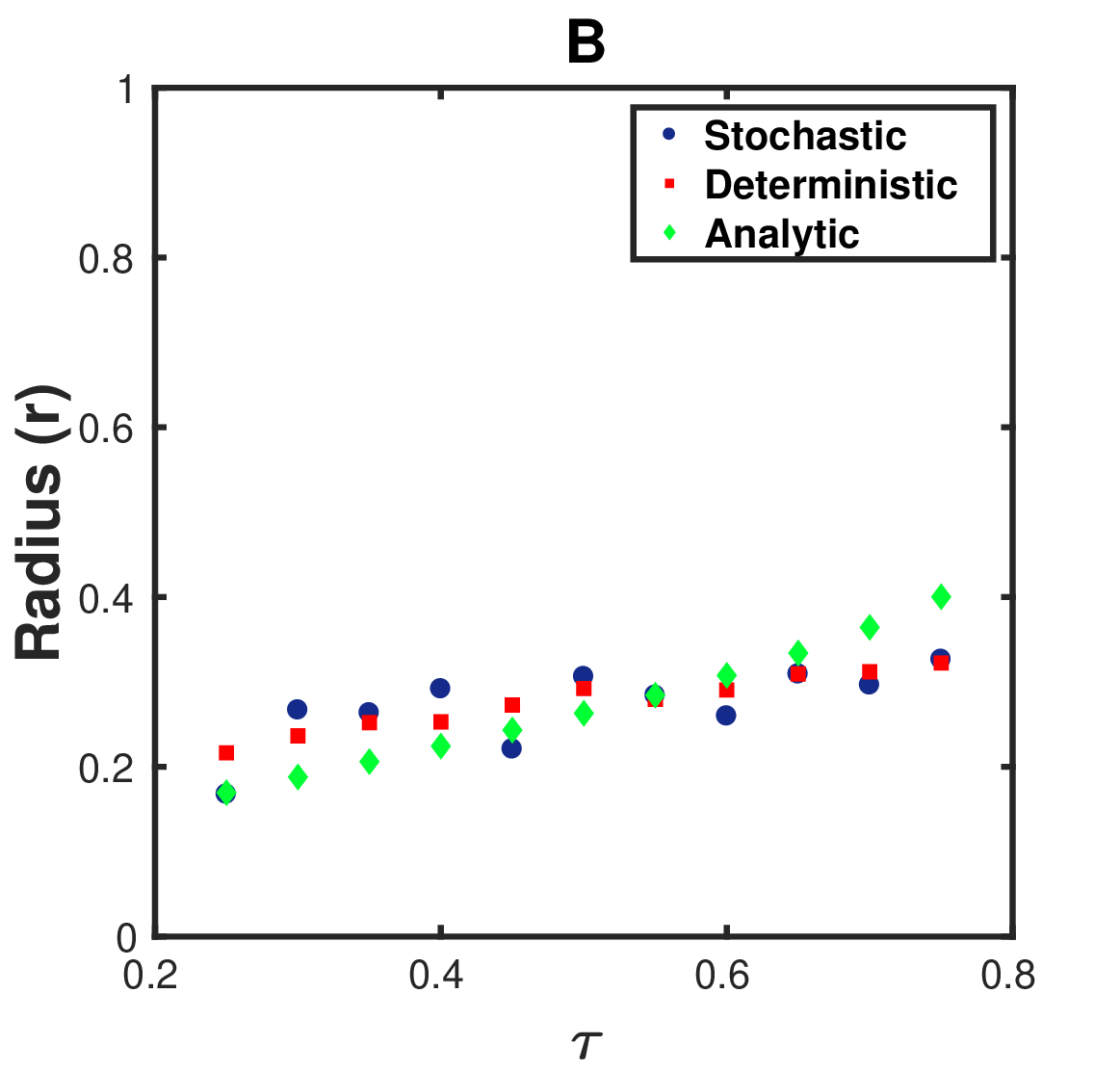}
\end{minipage}

\caption{Comparison of stochastic, deterministic, and analytical radius values versus different model parameters. Panel A shows the radius as a function of $\sigma$ for $\tau = 0.5$. Panel B shows the radius as a function of $\tau$ for $\gamma = 0.06$. In both panels $a = 1$ and $N=25000$.   }
\label{fig:4}
\end{figure}

 Figure~\ref{fig:4} illustrates the dependency of the mean radius ($\left\langle r \right\rangle$) versus system parameters $\tau$ and $\sigma$ obtained using stochastic simulation and analytical approximation. One can see that, as expected, our approximation matches the simulations when the limit cycle trajectory remains in the vicinity of the fixed point $M_{3}$, i.e. for small values of $\gamma$. 
 As the amplitude of the limit cycle increases (i.e., when we increase the value of $\gamma$) the approximation become less and less accurate. This mismatch occurs because the shape of the limit cycle becomes more and more different from an ellipse.

\subsection{\label{sec:III-B}Angular Brownian Motion and The Noise Strength}

In addition to producing amplitude fluctuations, demographic stochasticity also determines fluctuations in the phase of these oscillations. In this section we focus on the latter. In that context, it is natural to approximate the phase trajectory as an angular Brownian motion~\cite{gardiner1985handbook}, with fixed frequency (corresponding to the one of eq.~\ref{eq6}), and an angular diffusion term, with diffusion constant $D$.
This ansatz is consistent with the results of figure~\ref{fig:2}, where the time evolution of the phase oscillates around a constant frequency. Therefore, the corresponding effective Langevin equation~\cite{gardiner1985handbook} for the angular position $\theta$ can be written as follows:
\begin{eqnarray}
 \dot{\theta} = \omega + \sqrt{D} \xi(t) \ , 
%&& \sigma^{2}_{\theta} = Dt \nonumber
\label{eq8}
\end{eqnarray}
where $D$ is the diffusion constant and $\xi(t)$ is a white, delta correlated, Gaussian noise.

\begin{figure}
\begin{minipage}{.49\hsize}
\includegraphics[width=\linewidth]{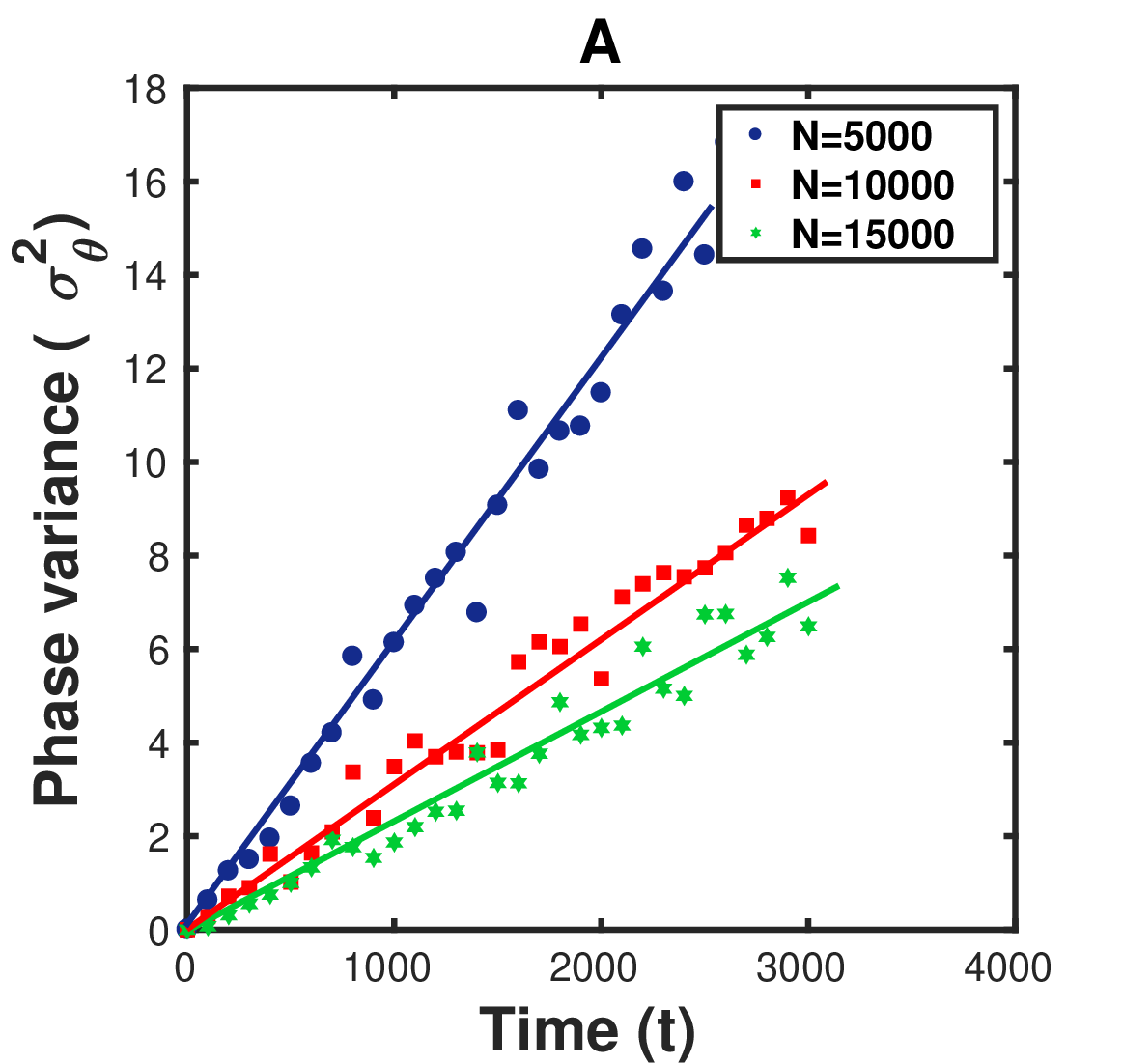}
\end{minipage}
\begin{minipage}{.49\hsize}
\includegraphics[width=\linewidth]{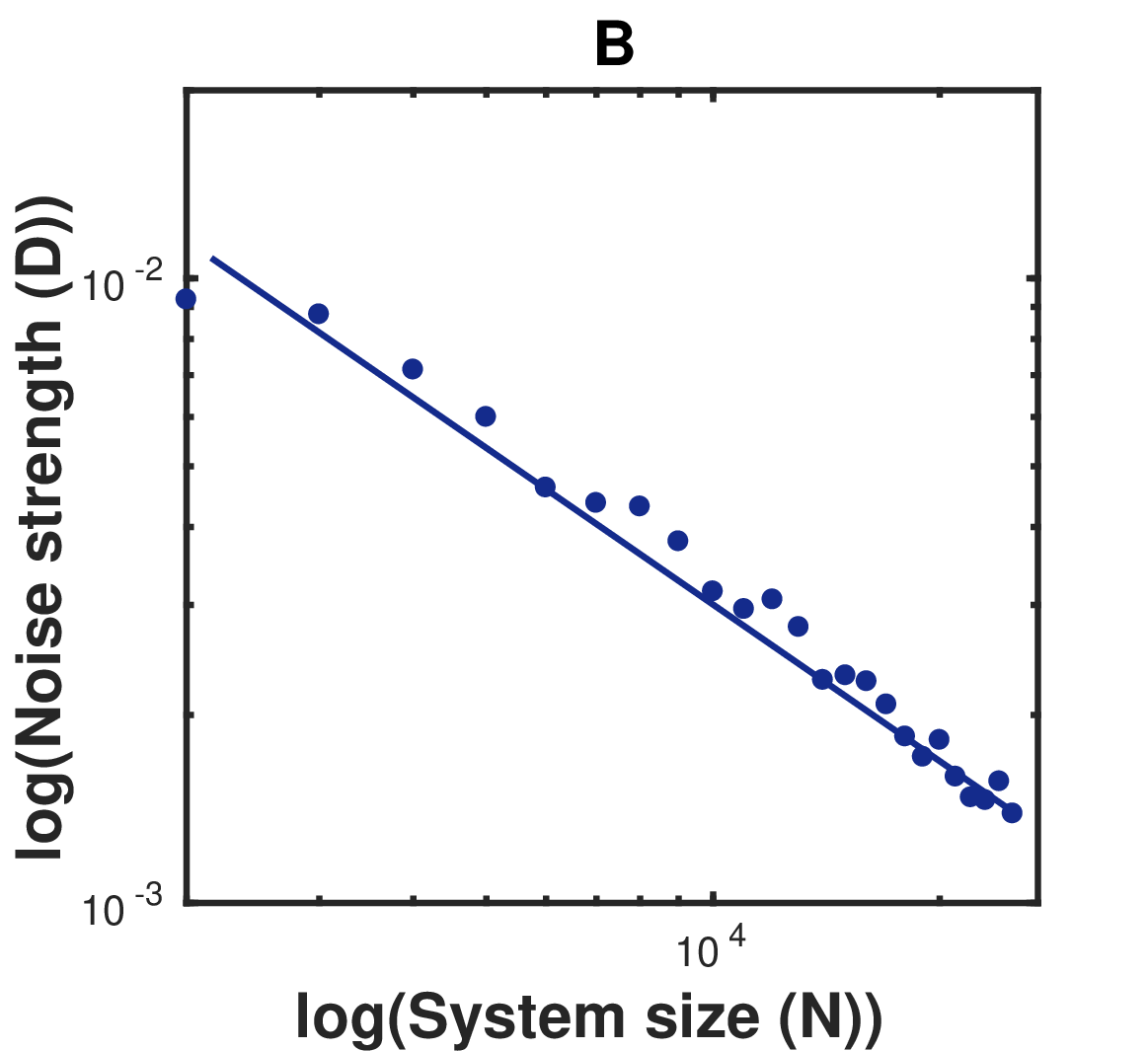}
\end{minipage}
\caption{Quantifying the noise strength along the limit cycle. (A) Phase variance over time for different values of system size. (B) the noise strength versus $N$. Parameters in these figures are $\tau = 0.5$, $a = 1$ and $\sigma = 3.05$ and the variance is calculated over 500 realizations of the system.}
\label{fig:5}
\end{figure}

Eq.~\ref{eq8} predicts that the variance of $\theta$ (calculated across independent realizations of the process starting from the same initial conditions) grows at $D t$.
Figure~\ref{fig:5} shows that, in agreement with the Geometric Brownian Motion ansatz, the variance of the phase increase linearly with time. 

The diffusion constant $D$ plays the role of an effective parameter, capturing the effect of stochasticity on the limit cycle trajectories. Our goal is to determine, in the linear noise approximation, the dependency of this effective diffusivity $D$ on the parameters of the RMA model ($a$, $\sigma$, and $\tau$) as well as on the population size $N$.

Given the demographic nature of the noise, we expect fluctuations to become less and less important with increasing $N$.
Figure~\ref{fig:5} shows that, as expected, $D$ has a reversed relationship with the system size $N$.

\subsection{\label{sec:III-C}Analytic Approximation for the Noise Strength}

In order to study the perturbations during evolution of the system around the limit cycle, we write the Fokker-Planck equation~\cite{gardiner1985handbook} of the stochastic RMA model by expanding the master equation for large values of $N$. We obtain the following equation:
\begin{eqnarray}
\begin{split}
\partial_{t}P(x,y) = & -\dfrac{\partial}{\partial x} F_{x}P(x,y) - \dfrac{\partial}{\partial y}F_{y}P(x,y)\\
& + \dfrac{1}{2N}\dfrac{\partial^{2}}{\partial x^{2}} B_{xx}P(x,y) + \dfrac{1}{2N}\dfrac{\partial^{2}}{\partial y^{2}}B_{yy} P(x,y) \\
& + \dfrac{1}{2N}\dfrac{\partial^{2}}{\partial x \partial y}B_{xy} P(x,y) + \dfrac{1}{2N}\dfrac{\partial^{2}}{\partial y \partial x}B_{yx} P(x,y), 
\end{split}
\label{eq18}
\end{eqnarray}
where $P(x,y)$, is the probability to find the system with abundances $(x,y) = (R/N,F/N)$ at time $t$ and:
\begin{eqnarray}
&& F_{x} = ax-\dfrac{x^{2}}{2}-\dfrac{\sigma xy}{1+\sigma \tau x}, \nonumber \\
&& F_{y} = \dfrac{\sigma xy}{1+\sigma \tau x} -y,
\nonumber \\
&& B_{xx} = \dfrac{1}{N}(ax + \dfrac{x^{2}}{2} + \dfrac{\sigma xy}{1+\sigma \tau x}), \nonumber \\
&& B_{yy} = \dfrac{1}{N}(\dfrac{\sigma xy}{1+\sigma \tau x} + y), \nonumber \\
&& B_{xy} = B_{yx} = - \dfrac{1}{N}(\dfrac{\sigma xy}{1+\sigma \tau x}).
\end{eqnarray}

In the limit of weak noise, the Fokker-Planck equation can be expanded around the deterministic solution, scaled by the square root of the noise amplitude ($\mathbf{v} = \dfrac{1}{\sqrt{N}}(\mathbf{x}-\mathbf{x_{det}})$). This is called Van Kampen expansion~\cite{van1992stochastic, gardiner1985handbook} and it has been previously applied in the context of ecological models with demographic stochasticity~\cite{BLACK2012337}. Under the Van Kampen expansion, by omitting terms of higher order in $(1/N)^{-1/2}$, the Fokker-Planck equation  can be transformed to the following equation:
\begin{eqnarray}
\begin{split}
\partial_{t}P(\mathbf{v},t)= & -\sum_{i,j}A_{ij}(t)\dfrac{\partial}{\partial v_{i}}v_{i}P(\mathbf{v},t)  \\
& +\dfrac{1}{2}\sum_{i,j}\tilde{B}_{ij}\dfrac{\partial^{2}}{\partial v_{i} \partial v_{j}}P(\mathbf{v},t),
\end{split}
\label{eq19}
\end{eqnarray}
where
\begin{eqnarray}
&& A_{ij} = \dfrac{\partial F_{i}}{\partial v_{j}}\vert_{\mathbf{x_{det}(t)}}, \nonumber \\
&& \tilde{B}_{ij} = B_{ij}\vert_{\mathbf{x_{det}(t)}}.
\label{eq20}
\end{eqnarray}

Since we approximated the shape of the limit cycle with an ellipse, we can write the deterministic solution in the following form:
\begin{eqnarray}
&& \mathbf{x_{det}} = \mathbf{x}_{f} + \begin{bmatrix}
r(t)\cos\theta(t)\\
\epsilon r(t) \sin\theta(t) \
\end{bmatrix}.
\label{eq45}
\end{eqnarray}

To investigate the stochastic RMA system, we decompose the noise term into the normal $\xi_n$ and tangent $\xi_l$ direction along the limit cycle, corresponding respectively to amplitude and phase fluctuations. The two components $\xi_n$ and $\xi_l$ can be obtained using:
\begin{eqnarray}
&& \mathbf{x} = \mathbf{x_{det}} + (r^{*}(t)+ \dfrac{\xi_{n}}{\sqrt{N}})\begin{bmatrix}
\cos(\omega t +\dfrac{\xi_{l}}{\sqrt{N}})\\
\epsilon \sin(\omega t +\dfrac{\xi_{l}}{\sqrt{N}}) \
\end{bmatrix} \ ,
\label{eq46}
\end{eqnarray}
in this equations $r^{*}=\left\langle r \right\rangle  $ and $\omega=\epsilon$ according to the eq.~\ref{eq6}. By expanding the noise terms for large $N$ we find that:
\begin{eqnarray}
&& \mathbf{x_{det}} = \mathbf{x_{f}}+ \begin{bmatrix}
r^{*}(t)\cos(\omega t)\\
\epsilon r^{*}(t)\sin(\omega t)
\end{bmatrix} +(\dfrac{1}{\sqrt{N}}) \begin{bmatrix}
\xi_{x}\\
\xi_{y}
\end{bmatrix} \ ,
\label{eq47}
\end{eqnarray}
where
\begin{eqnarray}
&& \xi_{x} = \xi_{n}\cos(\omega t) - \xi_{l}r^{*}(t)\sin(\omega t), \nonumber \\
&& \xi_{y} = \xi_{n}\epsilon \sin(\omega t) - \xi_{l}\epsilon r^{*}(t)\cos(\omega t) \ .
\label{eq48}
\end{eqnarray}

The Langevin equation corresponding to the $\xi_{x}$ and $\xi_{y}$  are written as:
\begin{eqnarray}
&& \dot{\xi_{x}} = \sum_{i} A_{x,i} + \eta_{x}, \nonumber \\
&& \dot{\xi_{y}} = \sum_{i} A_{y,i} + \eta_{y},
\label{eq50}
\end{eqnarray} 
where $\eta$ is related to the noise terms of Fokker-Planck equations as following:
\begin{eqnarray} 
&& \left\langle \eta_{x}^{2} \right\rangle = B_{xx}, \nonumber \\
&& \left\langle \eta_{y}^{2}\right\rangle = B_{yy},
\nonumber \\
&& \left\langle \eta_{x} \eta_{y}\right\rangle = B_{xy} \ .
\label{eq53}
\end{eqnarray}

\begin{figure*}
%\begin{minipage}[t]{0.3\textwidth}
 % \includegraphics[width=\linewidth]{SIX-A-N.eps}
  %\caption{First}
 % \label{fig:first}
%\end{minipage}%
%\hfill % maximize the horizontal separation
\begin{minipage}[t]{0.49\textwidth}
  \includegraphics[width=\linewidth]{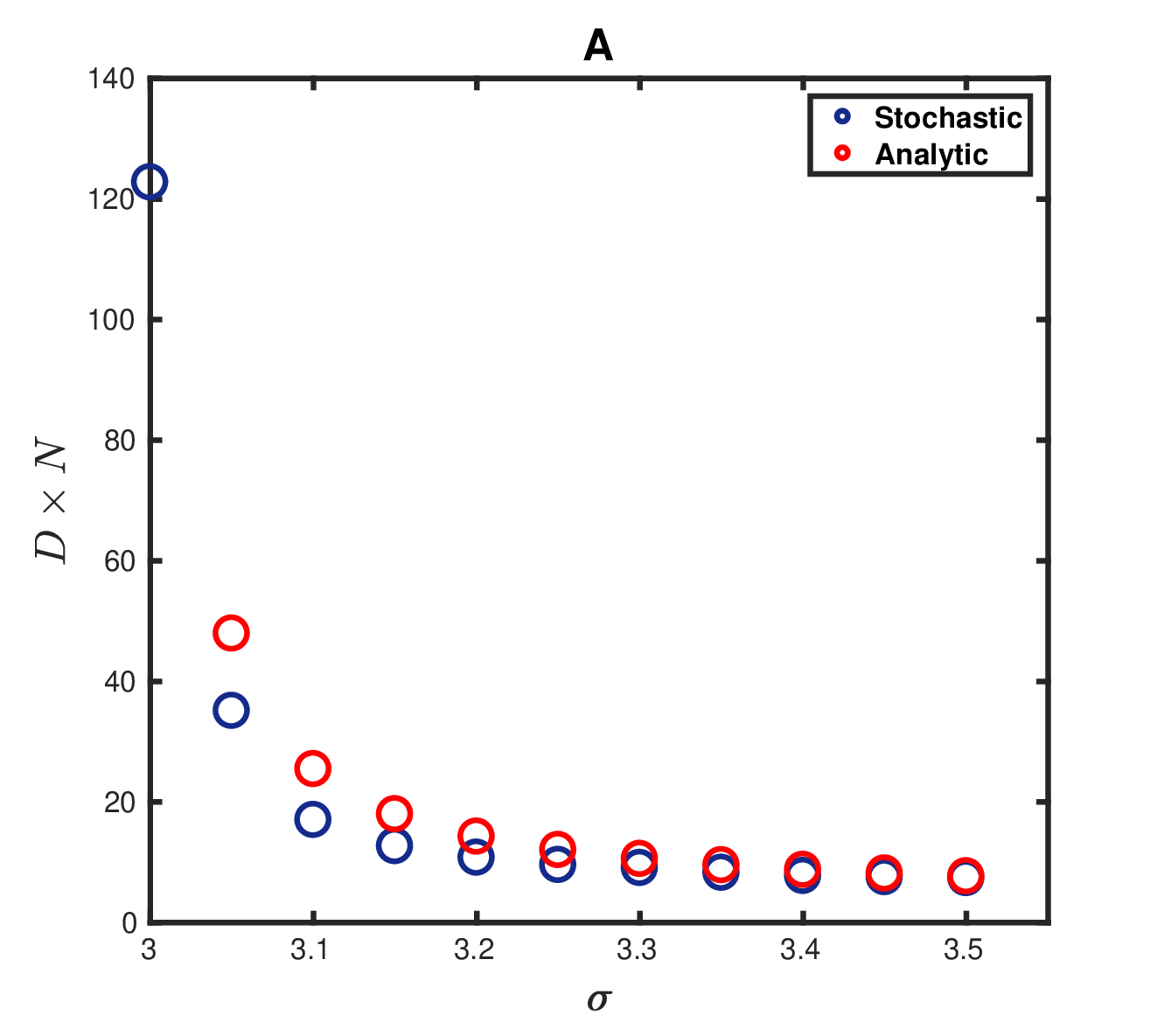}
  %\caption{Second}
  \label{fig:second}
\end{minipage}%
\hfill
\begin{minipage}[t]{0.49\textwidth}
  \includegraphics[width=\linewidth]{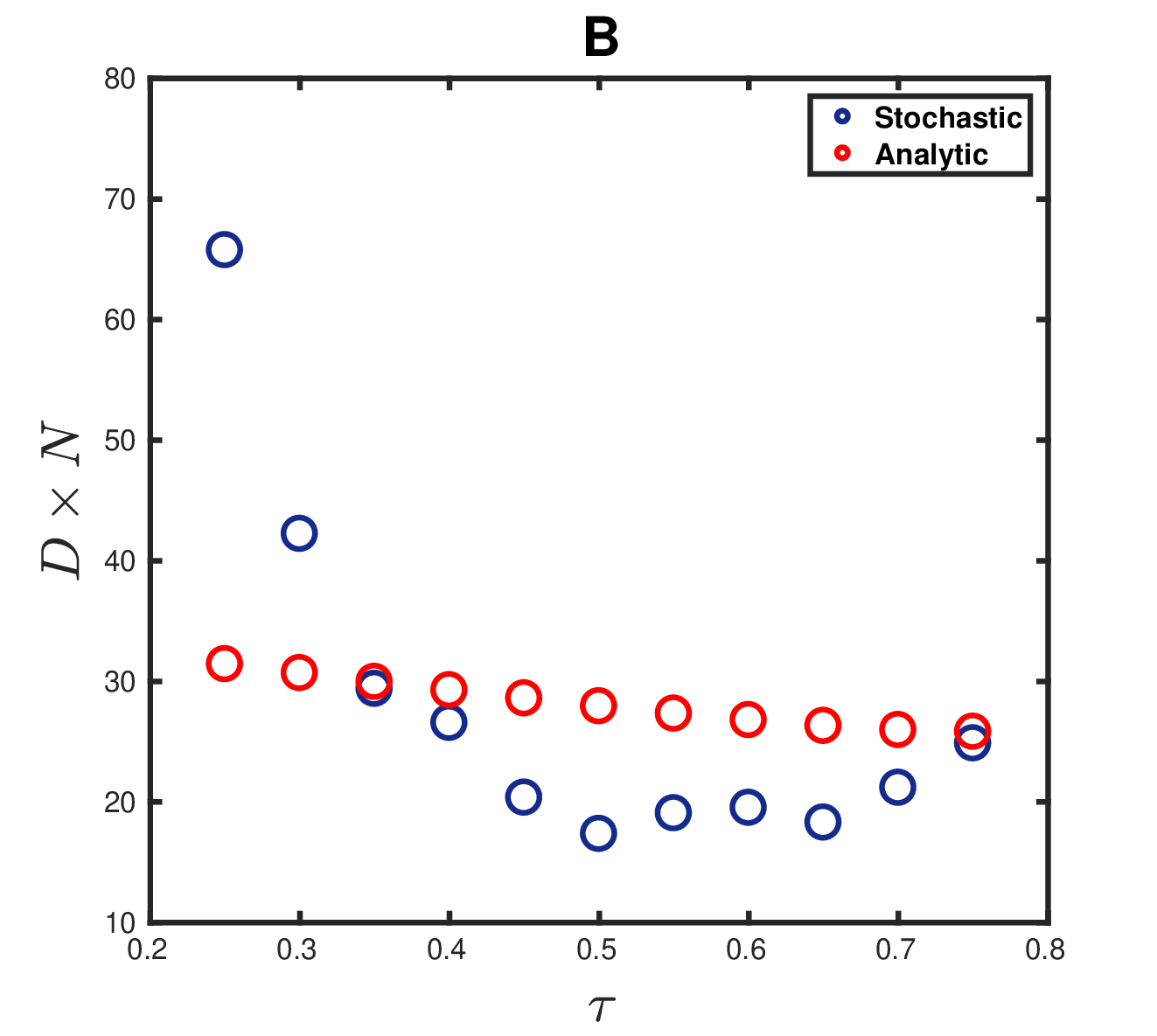}
  %\caption{Third}
  \label{fig:third}
\end{minipage}%
\caption{Comparison of the stochastic and analytical noise strength for different parameters. Parameters are $\gamma = 0.03$, $a = 1$, $\tau = 0.5$ and $N=25000$.}
\label{fig:6}
\end{figure*}

The value of $\left\langle \dot{\xi_{l}^{2}} \right\rangle $ can be found by straight forward calculations from equations~\ref{eq48} and~\ref{eq50}as follows  
\begin{eqnarray}
&& \left\langle \dot{\xi_{l}^{2}} \right\rangle = \left\langle \dot{\xi_{l}^{2}} \right\rangle_{det}+Dt, \\
&& D = \dfrac{1}{2r^{2}\epsilon^{2}N}\left[ \epsilon^{2}(ax + \dfrac{x^{2}}{2} + \dfrac{xy\sigma}{1+x\tau\sigma})+ y+\dfrac{xy\sigma}{1+x\tau\sigma}\right] \ . \nonumber 
\label{eq56}
\end{eqnarray}
Where the averaging is over different realizations and one period of time. Figure~\ref{fig:5} shows the noise strength inferred from stochastic simulations, decreases as $1/\sqrt{N}$ as predicted by our analytical calculation. Figure~\ref{fig:6} shows the dependency of the effective diffusion constant $D$ on the parameters $\sigma$ and $\tau$. In the regime where our approximation is expected to work, i.e., when $\sigma$ is close to the $\sigma^{*}$ (i.e., $\gamma \to 0$) the results of analytical approximation matches well the numerical simulations. 

%\lipsum[1] 
%\lipsum[2-19]

\section{\label{sec:IV} Discussion}

Oscillatory behavior in the populations has been widely reported in natural ecosystems and laboratory-scale experiments. Mathematical modeling of the interaction between predator and prey populations leads to the nonlinear ordinary differential equations that the time-dependent solutions of them are associated with  limit cycle oscillations.  Demographic stochasticity, due to unpredictability of the timing of birth, death, and interaction events, is often neglected and can remarkably alter the dynamics of a system. 

In this paper, we have quantified the strength of the perturbations, due to the demographic effects of interacting species, in the limit cycle of the  Roseznweig-MacArthur model.  We showed that, in the vicinity of the bifurcation point, the limit cycle shape is well approximated by an ellipse. In this limit, we obtained analytically the mean frequency and mean radius of the limit cycle as functions of the model parameters.

We then focus on the effect of stochasticity on the fluctuations along the limit cycle, that we effectively model as an angular Brownian motion. By expanding the master equation of the population densities through the Van Kampen system size method, we found demographic noise strength analytically, which has an inverse relationship with the system size.
 Finally, we compared our analytical results with the results found using simulation of the stochastic Rosenzweig-MacArthur model by Gillespie algorithm. A close agreement is observed between analytical results and the results found from the simulations. 

Our work shows how the effect of demographic stochasticity on the frequency and phase of oscillation can be analytically captured even for non-linear model. Our calculation can be potentially extended to consider different sources of stochasticity (e.g. environmental) or to analyses oscillations in spatially extended systems.

\begin{acknowledgments}
SG acknowledges the ICTP STEP program for the support for her visits to The Abdus Salam ICTP.
\end{acknowledgments}

% The \nocite command causes all entries in a bibliography to be printed out
% whether or not they are actually referenced in the text. This is appropriate
% for the sample file to show the different styles of references, but authors
% most likely will not want to use it.
%\nocite{*}

\bibliography{main2}% Produces the bibliography via BibTeX.

%apsrev4-2.bst 2019-01-14 (MD) hand-edited version of apsrev4-1.bst
%Control: key (0)
%Control: author (8) initials jnrlst
%Control: editor formatted (1) identically to author
%Control: production of article title (0) allowed
%Control: page (0) single
%Control: year (1) truncated
%Control: production of eprint (0) enabled
\providecommand{\noopsort}[1]{}\providecommand{\singleletter}[1]{#1}%
\begin{thebibliography}{20}%
\makeatletter
\providecommand \@ifxundefined [1]{%
 \@ifx{#1\undefined}
}%
\providecommand \@ifnum [1]{%
 \ifnum #1\expandafter \@firstoftwo
 \else \expandafter \@secondoftwo
 \fi
}%
\providecommand \@ifx [1]{%
 \ifx #1\expandafter \@firstoftwo
 \else \expandafter \@secondoftwo
 \fi
}%
\providecommand \natexlab [1]{#1}%
\providecommand \enquote  [1]{``#1''}%
\providecommand \bibnamefont  [1]{#1}%
\providecommand \bibfnamefont [1]{#1}%
\providecommand \citenamefont [1]{#1}%
\providecommand \href@noop [0]{\@secondoftwo}%
\providecommand \href [0]{\begingroup \@sanitize@url \@href}%
\providecommand \@href[1]{\@@startlink{#1}\@@href}%
\providecommand \@@href[1]{\endgroup#1\@@endlink}%
\providecommand \@sanitize@url [0]{\catcode `\\12\catcode `\$12\catcode
  `\&12\catcode `\#12\catcode `\^12\catcode `\_12\catcode `\%12\relax}%
\providecommand \@@startlink[1]{}%
\providecommand \@@endlink[0]{}%
\providecommand \url  [0]{\begingroup\@sanitize@url \@url }%
\providecommand \@url [1]{\endgroup\@href {#1}{\urlprefix }}%
\providecommand \urlprefix  [0]{URL }%
\providecommand \Eprint [0]{\href }%
\providecommand \doibase [0]{https://doi.org/}%
\providecommand \selectlanguage [0]{\@gobble}%
\providecommand \bibinfo  [0]{\@secondoftwo}%
\providecommand \bibfield  [0]{\@secondoftwo}%
\providecommand \translation [1]{[#1]}%
\providecommand \BibitemOpen [0]{}%
\providecommand \bibitemStop [0]{}%
\providecommand \bibitemNoStop [0]{.\EOS\space}%
\providecommand \EOS [0]{\spacefactor3000\relax}%
\providecommand \BibitemShut  [1]{\csname bibitem#1\endcsname}%
\let\auto@bib@innerbib\@empty
%</preamble>
\bibitem [{\citenamefont {Lotka}(1920)}]{lotka1920analytical}%
  \BibitemOpen
  \bibfield  {author} {\bibinfo {author} {\bibfnamefont {A.~J.}\ \bibnamefont
  {Lotka}},\ }\bibfield  {title} {\bibinfo {title} {Analytical note on certain
  rhythmic relations in organic systems},\ }\href@noop {} {\bibfield  {journal}
  {\bibinfo  {journal} {Proceedings of the National Academy of Sciences}\
  }\textbf {\bibinfo {volume} {6}},\ \bibinfo {pages} {410} (\bibinfo {year}
  {1920})}\BibitemShut {NoStop}%
\bibitem [{\citenamefont {Volterra}(1927)}]{volterra1927variazioni}%
  \BibitemOpen
  \bibfield  {author} {\bibinfo {author} {\bibfnamefont {V.}~\bibnamefont
  {Volterra}},\ }\href@noop {} {\emph {\bibinfo {title} {Variazioni e
  fluttuazioni del numero d'individui in specie animali conviventi}}},\
  Vol.~\bibinfo {volume} {2}\ (\bibinfo  {publisher} {Societ{\'a} anonima
  tipografica" Leonardo da Vinci"},\ \bibinfo {year} {1927})\BibitemShut
  {NoStop}%
\bibitem [{\citenamefont {Rosenzweig}\ and\ \citenamefont
  {MacArthur}(1963)}]{rosenzweig1963graphical}%
  \BibitemOpen
  \bibfield  {author} {\bibinfo {author} {\bibfnamefont {M.~L.}\ \bibnamefont
  {Rosenzweig}}\ and\ \bibinfo {author} {\bibfnamefont {R.~H.}\ \bibnamefont
  {MacArthur}},\ }\bibfield  {title} {\bibinfo {title} {Graphical
  representation and stability conditions of predator-prey interactions},\
  }\href@noop {} {\bibfield  {journal} {\bibinfo  {journal} {The American
  Naturalist}\ }\textbf {\bibinfo {volume} {97}},\ \bibinfo {pages} {209}
  (\bibinfo {year} {1963})}\BibitemShut {NoStop}%
\bibitem [{\citenamefont {Holling}(1959)}]{holling1959components}%
  \BibitemOpen
  \bibfield  {author} {\bibinfo {author} {\bibfnamefont {C.~S.}\ \bibnamefont
  {Holling}},\ }\bibfield  {title} {\bibinfo {title} {The components of
  predation as revealed by a study of small-mammal predation of the european
  pine sawfly1},\ }\href@noop {} {\bibfield  {journal} {\bibinfo  {journal}
  {The canadian entomologist}\ }\textbf {\bibinfo {volume} {91}},\ \bibinfo
  {pages} {293} (\bibinfo {year} {1959})}\BibitemShut {NoStop}%
\bibitem [{\citenamefont {Gilpin}(1973)}]{gilpin1973hares}%
  \BibitemOpen
  \bibfield  {author} {\bibinfo {author} {\bibfnamefont {M.~E.}\ \bibnamefont
  {Gilpin}},\ }\bibfield  {title} {\bibinfo {title} {Do hares eat lynx?},\
  }\href@noop {} {\bibfield  {journal} {\bibinfo  {journal} {The American
  Naturalist}\ }\textbf {\bibinfo {volume} {107}},\ \bibinfo {pages} {727}
  (\bibinfo {year} {1973})}\BibitemShut {NoStop}%
\bibitem [{\citenamefont {Elton}\ and\ \citenamefont
  {Nicholson}(1942)}]{elton1942ten}%
  \BibitemOpen
  \bibfield  {author} {\bibinfo {author} {\bibfnamefont {C.}~\bibnamefont
  {Elton}}\ and\ \bibinfo {author} {\bibfnamefont {M.}~\bibnamefont
  {Nicholson}},\ }\bibfield  {title} {\bibinfo {title} {The ten-year cycle in
  numbers of the lynx in canada},\ }\href@noop {} {\bibfield  {journal}
  {\bibinfo  {journal} {The Journal of Animal Ecology}\ ,\ \bibinfo {pages}
  {215}} (\bibinfo {year} {1942})}\BibitemShut {NoStop}%
\bibitem [{\citenamefont {Butler}(1953)}]{butler1953nature}%
  \BibitemOpen
  \bibfield  {author} {\bibinfo {author} {\bibfnamefont {L.}~\bibnamefont
  {Butler}},\ }\bibfield  {title} {\bibinfo {title} {The nature of cycles in
  populations of canadian mammals},\ }\href@noop {} {\bibfield  {journal}
  {\bibinfo  {journal} {Canadian Journal of Zoology}\ }\textbf {\bibinfo
  {volume} {31}},\ \bibinfo {pages} {242} (\bibinfo {year} {1953})}\BibitemShut
  {NoStop}%
\bibitem [{\citenamefont {Korpim{\"a}ki}\ \emph {et~al.}(2005)\citenamefont
  {Korpim{\"a}ki}, \citenamefont {Norrdahl}, \citenamefont {Huitu},\ and\
  \citenamefont {Klemola}}]{korpimaki2005predator}%
  \BibitemOpen
  \bibfield  {author} {\bibinfo {author} {\bibfnamefont {E.}~\bibnamefont
  {Korpim{\"a}ki}}, \bibinfo {author} {\bibfnamefont {K.}~\bibnamefont
  {Norrdahl}}, \bibinfo {author} {\bibfnamefont {O.}~\bibnamefont {Huitu}},\
  and\ \bibinfo {author} {\bibfnamefont {T.}~\bibnamefont {Klemola}},\
  }\bibfield  {title} {\bibinfo {title} {Predator--induced synchrony in
  population oscillations of coexisting small mammal species},\ }\href@noop {}
  {\bibfield  {journal} {\bibinfo  {journal} {Proceedings of the Royal Society
  B: Biological Sciences}\ }\textbf {\bibinfo {volume} {272}},\ \bibinfo
  {pages} {193} (\bibinfo {year} {2005})}\BibitemShut {NoStop}%
\bibitem [{\citenamefont {Higgins}\ \emph {et~al.}(1997)\citenamefont
  {Higgins}, \citenamefont {Hastings}, \citenamefont {Sarvela},\ and\
  \citenamefont {Botsford}}]{higgins1997stochastic}%
  \BibitemOpen
  \bibfield  {author} {\bibinfo {author} {\bibfnamefont {K.}~\bibnamefont
  {Higgins}}, \bibinfo {author} {\bibfnamefont {A.}~\bibnamefont {Hastings}},
  \bibinfo {author} {\bibfnamefont {J.~N.}\ \bibnamefont {Sarvela}},\ and\
  \bibinfo {author} {\bibfnamefont {L.~W.}\ \bibnamefont {Botsford}},\
  }\bibfield  {title} {\bibinfo {title} {Stochastic dynamics and deterministic
  skeletons: population behavior of dungeness crab},\ }\href@noop {} {\bibfield
   {journal} {\bibinfo  {journal} {Science}\ }\textbf {\bibinfo {volume}
  {276}},\ \bibinfo {pages} {1431} (\bibinfo {year} {1997})}\BibitemShut
  {NoStop}%
\bibitem [{\citenamefont {Kamata}\ and\ \citenamefont
  {Liebhold}(2000)}]{kamata2000population}%
  \BibitemOpen
  \bibfield  {author} {\bibinfo {author} {\bibfnamefont {N.}~\bibnamefont
  {Kamata}}\ and\ \bibinfo {author} {\bibfnamefont {A.}~\bibnamefont
  {Liebhold}},\ }\bibfield  {title} {\bibinfo {title} {Are population cycles
  and spatial synchrony a universal characteristic of forest insect
  populations?},\ }\href@noop {} {\bibfield  {journal} {\bibinfo  {journal}
  {Population Ecology}\ }\textbf {\bibinfo {volume} {42}} (\bibinfo {year}
  {2000})}\BibitemShut {NoStop}%
\bibitem [{\citenamefont {May}(1973)}]{may1973stability}%
  \BibitemOpen
  \bibfield  {author} {\bibinfo {author} {\bibfnamefont {R.~M.}\ \bibnamefont
  {May}},\ }\bibfield  {title} {\bibinfo {title} {Stability in randomly
  fluctuating versus deterministic environments},\ }\href@noop {} {\bibfield
  {journal} {\bibinfo  {journal} {The American Naturalist}\ }\textbf {\bibinfo
  {volume} {107}},\ \bibinfo {pages} {621} (\bibinfo {year}
  {1973})}\BibitemShut {NoStop}%
\bibitem [{\citenamefont {Melbourne}\ and\ \citenamefont
  {Hastings}(2008)}]{melbourne2008extinction}%
  \BibitemOpen
  \bibfield  {author} {\bibinfo {author} {\bibfnamefont {B.~A.}\ \bibnamefont
  {Melbourne}}\ and\ \bibinfo {author} {\bibfnamefont {A.}~\bibnamefont
  {Hastings}},\ }\bibfield  {title} {\bibinfo {title} {Extinction risk depends
  strongly on factors contributing to stochasticity},\ }\href@noop {}
  {\bibfield  {journal} {\bibinfo  {journal} {Nature}\ }\textbf {\bibinfo
  {volume} {454}},\ \bibinfo {pages} {100} (\bibinfo {year}
  {2008})}\BibitemShut {NoStop}%
\bibitem [{\citenamefont {Petrovskii}\ \emph {et~al.}(2010)\citenamefont
  {Petrovskii}, \citenamefont {Morozov}, \citenamefont {Malchow},\ and\
  \citenamefont {Sieber}}]{Petrovskii2010}%
  \BibitemOpen
  \bibfield  {author} {\bibinfo {author} {\bibfnamefont {S.}~\bibnamefont
  {Petrovskii}}, \bibinfo {author} {\bibfnamefont {A.}~\bibnamefont {Morozov}},
  \bibinfo {author} {\bibfnamefont {H.}~\bibnamefont {Malchow}},\ and\ \bibinfo
  {author} {\bibfnamefont {M.}~\bibnamefont {Sieber}},\ }\bibfield  {title}
  {\bibinfo {title} {Noise can prevent onset of chaos in spatiotemporal
  population dynamics},\ }\href {https://doi.org/10.1140/epjb/e2010-10410-8}
  {\bibfield  {journal} {\bibinfo  {journal} {The European Physical Journal B}\
  }\textbf {\bibinfo {volume} {78}},\ \bibinfo {pages} {253} (\bibinfo {year}
  {2010})}\BibitemShut {NoStop}%
\bibitem [{\citenamefont {Reuman}\ \emph {et~al.}(2008)\citenamefont {Reuman},
  \citenamefont {Costantino}, \citenamefont {Desharnais},\ and\ \citenamefont
  {Cohen}}]{reuman2008colour}%
  \BibitemOpen
  \bibfield  {author} {\bibinfo {author} {\bibfnamefont {D.~C.}\ \bibnamefont
  {Reuman}}, \bibinfo {author} {\bibfnamefont {R.~F.}\ \bibnamefont
  {Costantino}}, \bibinfo {author} {\bibfnamefont {R.~A.}\ \bibnamefont
  {Desharnais}},\ and\ \bibinfo {author} {\bibfnamefont {J.~E.}\ \bibnamefont
  {Cohen}},\ }\bibfield  {title} {\bibinfo {title} {Colour of environmental
  noise affects the nonlinear dynamics of cycling, stage-structured
  populations},\ }\href@noop {} {\bibfield  {journal} {\bibinfo  {journal}
  {Ecology Letters}\ }\textbf {\bibinfo {volume} {11}},\ \bibinfo {pages} {820}
  (\bibinfo {year} {2008})}\BibitemShut {NoStop}%
\bibitem [{\citenamefont {Black}\ and\ \citenamefont
  {McKane}(2010)}]{black2010stochastic}%
  \BibitemOpen
  \bibfield  {author} {\bibinfo {author} {\bibfnamefont {A.~J.}\ \bibnamefont
  {Black}}\ and\ \bibinfo {author} {\bibfnamefont {A.~J.}\ \bibnamefont
  {McKane}},\ }\bibfield  {title} {\bibinfo {title} {Stochastic amplification
  in an epidemic model with seasonal forcing},\ }\href@noop {} {\bibfield
  {journal} {\bibinfo  {journal} {Journal of Theoretical Biology}\ }\textbf
  {\bibinfo {volume} {267}},\ \bibinfo {pages} {85} (\bibinfo {year}
  {2010})}\BibitemShut {NoStop}%
\bibitem [{\citenamefont {Cheng}(1981)}]{doi:10.1137/0512047}%
  \BibitemOpen
  \bibfield  {author} {\bibinfo {author} {\bibfnamefont {K.-S.}\ \bibnamefont
  {Cheng}},\ }\bibfield  {title} {\bibinfo {title} {Uniqueness of a limit cycle
  for a predator-prey system},\ }\href {https://doi.org/10.1137/0512047}
  {\bibfield  {journal} {\bibinfo  {journal} {SIAM Journal on Mathematical
  Analysis}\ }\textbf {\bibinfo {volume} {12}},\ \bibinfo {pages} {541}
  (\bibinfo {year} {1981})},\ \Eprint
  {https://arxiv.org/abs/https://doi.org/10.1137/0512047}
  {https://doi.org/10.1137/0512047} \BibitemShut {NoStop}%
\bibitem [{\citenamefont {Smith}\ and\ \citenamefont
  {Meerson}(2016)}]{smith2016extinction}%
  \BibitemOpen
  \bibfield  {author} {\bibinfo {author} {\bibfnamefont {N.~R.}\ \bibnamefont
  {Smith}}\ and\ \bibinfo {author} {\bibfnamefont {B.}~\bibnamefont
  {Meerson}},\ }\bibfield  {title} {\bibinfo {title} {Extinction of oscillating
  populations},\ }\href@noop {} {\bibfield  {journal} {\bibinfo  {journal}
  {Physical Review E}\ }\textbf {\bibinfo {volume} {93}},\ \bibinfo {pages}
  {032109} (\bibinfo {year} {2016})}\BibitemShut {NoStop}%
\bibitem [{\citenamefont {Gardiner}\ \emph {et~al.}(1985)\citenamefont
  {Gardiner} \emph {et~al.}}]{gardiner1985handbook}%
  \BibitemOpen
  \bibfield  {author} {\bibinfo {author} {\bibfnamefont {C.~W.}\ \bibnamefont
  {Gardiner}} \emph {et~al.},\ }\href@noop {} {\emph {\bibinfo {title}
  {Handbook of stochastic methods}}},\ Vol.~\bibinfo {volume} {3}\ (\bibinfo
  {publisher} {springer Berlin},\ \bibinfo {year} {1985})\BibitemShut {NoStop}%
\bibitem [{\citenamefont {Van~Kampen}(1992)}]{van1992stochastic}%
  \BibitemOpen
  \bibfield  {author} {\bibinfo {author} {\bibfnamefont {N.~G.}\ \bibnamefont
  {Van~Kampen}},\ }\href@noop {} {\emph {\bibinfo {title} {Stochastic processes
  in physics and chemistry}}},\ Vol.~\bibinfo {volume} {1}\ (\bibinfo
  {publisher} {Elsevier},\ \bibinfo {year} {1992})\BibitemShut {NoStop}%
\bibitem [{\citenamefont {Black}\ and\ \citenamefont
  {McKane}(2012)}]{BLACK2012337}%
  \BibitemOpen
  \bibfield  {author} {\bibinfo {author} {\bibfnamefont {A.~J.}\ \bibnamefont
  {Black}}\ and\ \bibinfo {author} {\bibfnamefont {A.~J.}\ \bibnamefont
  {McKane}},\ }\bibfield  {title} {\bibinfo {title} {Stochastic formulation of
  ecological models and their applications},\ }\href
  {https://doi.org/https://doi.org/10.1016/j.tree.2012.01.014} {\bibfield
  {journal} {\bibinfo  {journal} {Trends in Ecology \& Evolution}\ }\textbf
  {\bibinfo {volume} {27}},\ \bibinfo {pages} {337} (\bibinfo {year}
  {2012})}\BibitemShut {NoStop}%
\end{thebibliography}%

\end{document}

%\preprint{APS/123-QED}

\section{Supplementary material}% Force line breaks with \\
%\thanks{A footnote to the article title}%
Here we write the detailed derivation of equation 8 of the paper in subsection~\ref{ONE} and equation 15 in subsection~\ref{TWO}.
\subsection{\label{ONE}{Derivation of the radius and angular velocity of the stable limit cycle}}
To be able to find the approximate value of the radius and angular velocity along the limit cycle, one needs to first find the population density evolution equations in polar coordinates. For this purpose, we started expanding the equations close to the coexisting fixed point.
%in subsection~\ref{First}. 
Then, %in subsection~\ref{Second}, 
we supposed that the shape of the limit cycle close to the bifurcation is elliptical.
%The detailed derivations are explained in subsections~\ref{Third} and~\ref{Fourth} successively. 
%\subsection{\label{First}{Expanding Non-Linear Equations Around Fixed-Point}}

Initially, to find the equations for deviation from the coexisting fixed point of the system, we used Taylor expansion around the mentioned fixed point. Considering the main RMA density equations:
\begin{eqnarray} 
%\begin{split}
&& \dot{x} = ax - \dfrac{x^2}{2}- \dfrac{\sigma xy}{1+ \sigma \tau x},  \\ 
&& \dot{y} = -y + \dfrac{\sigma xy}{1+ \sigma \tau x},\nonumber,
%\end{split}
\label{eq22}
\end{eqnarray}
we use second order Taylor expansion of the above equations around the fixed point $(x_{0}, y_{0})$. Therefore, we will have the following equations:
\begin{align}
\dfrac{d(x-x_{0})}{dt} &= ax_{0} - \dfrac{x_{0}^2}{2}- \dfrac{\sigma x_{0}y_{0}}{1+ \sigma \tau x_{0}} - \dfrac{\sigma x_{0}(y-y_{0})}{1+ \sigma \tau x_{0}}        \\
& + \left[ a - x_{0} - \dfrac{y_{0}\sigma}{(1+ \sigma \tau x_{0})^2} - \dfrac{(y-y_{0})\sigma}{(1+ \sigma \tau x_{0})^2} \right]  (x-x_{0}) \nonumber \\
&  + \left[ -\dfrac{1}{2} + \dfrac{y_{0}\sigma ^2 \tau}{(1+ \sigma \tau x_{0})^3} + \dfrac{(y-y_{0})\sigma ^2 \tau }{(1+ \sigma \tau x_{0})^3} \right] (x-x_{0})^2 \nonumber 
\label{eq23}
\end{align}

\begin{align}
\dfrac{d(y-y_{0})}{dt} &= -y_{0} + \dfrac{\sigma x_{0}y_{0}}{1+ \sigma \tau x_{0}}  \\
& + \left[ -1 + \dfrac{\sigma x_{0}}{1+ \sigma \tau x_{0}} \right] (y-y_{0})\nonumber \\
&+ \left[ \dfrac{y_{0}\sigma}{(1+ \sigma \tau x_{0})^2} + \dfrac{(y-y_{0})\sigma}{(1+ \sigma \tau x_{0})^2} \right] (x-x_{0}) \nonumber \\
& + \left[ -\dfrac{y_{0}\sigma ^2 \tau}{(1+ \sigma \tau x_{0})^3} - \dfrac{(y-y_{0})\sigma ^2 \tau }{(1+ \sigma \tau x_{0})^3} \right] (x-x_{0})^2 \nonumber 
\label{eq24}
\end{align}

One can substitute $(x-x_{0})\rightarrow dx$, $(y-y_{0})\rightarrow dy$ and $(x_{0}, y_{0}) \rightarrow (\frac{1}{\sigma (1-\tau)}, \dfrac{2a\sigma (1-\tau)-1 }{2\sigma^2 (1-\tau)^2})$. Therefore, the above equations transform to the following ones:
\begin{align}
\dot{dx}&= -dy + dx(\dfrac{1}{2\sigma}+\dfrac{1}{\sigma(\tau -1)} + a\tau)\\
& -dxdy(\sigma(\tau - 1)^2) - dx^{2}dy(\sigma ^2 \tau (\tau -1)^3) \nonumber \\
&+ dx^{2}((1/2)(-1 + (1 + 2a\sigma(\tau - 1))(\tau - 1)\tau)). \nonumber
\label{eq25}
\end{align}

\begin{align}
\dot{dy}&= dx(a-\dfrac{1}{2\sigma}-a\sigma)+dxdy(\sigma(\tau -1)^2) \\
& +dx^2dy(\sigma ^2\tau(\tau -1)^3)\nonumber \\ 
& + dx^{2}((1/2)(1 + 2a\sigma(\tau - 1)(\tau - 1)\tau)). \nonumber
\label{eq26}
\end{align}

%\subsection{\label{Second}{Elliptical approximation for limit cycle}}
As we mentioned before, close to the bifurcation, one  can approximate the shape of the limit cycle with an ellipse with longer radius along the x-axis. Therefore, we can write:
\begin{eqnarray}
&& dx = r(t)\cos\theta(t)  \\
&& dy = r(t)\epsilon \sin\theta(t). \nonumber
\label{eq27}
\end{eqnarray}
By substituting equation~\ref{eq27} in equations~\ref{eq25} and~\ref{eq26}, we obtain:
\begin{align}
\dot{dx} &=  \dot{r}(t)\cos\theta(t) - r(t)\dot{\theta}(t)\sin\theta(t)  \\ 
& = -r(t)\epsilon \sin\theta(t) + r(t)\cos\theta(t)\left[\dfrac{1}{2\sigma}+\dfrac{1}{\sigma(\tau -1)}+ a\tau \right]\nonumber \\ & -r(t)^2 \epsilon \cos \theta(t) \sin \theta(t) \left[ \sigma (\tau -1)^2 \right] \nonumber \\
&-r(t)^3 \epsilon \cos^2\theta(t)\sin\theta(t) \left[ \sigma ^2\tau (\tau -1)^3 \right]\nonumber \\ 
& + r(t)^{2} \cos^2\theta(t) \left[ (1/2)(-1 + (1 + 2a\sigma(\tau - 1))(\tau - 1)\tau)\right], \nonumber \qedhere
\label{eq28}
\end{align}

\begin{align}
\dot{dy} &= \dot{r}(t)\epsilon \sin \theta(t) + r(t)\epsilon \dot{\theta}(t)\cos\theta(t)   \\
&= r(t)\cos\theta(t)\left[a-\dfrac{1}{2\sigma}-a\sigma \right]\nonumber \\ &+ r(t)^2 \epsilon \cos \theta(t) \sin \theta(t) \left[ \sigma (\tau -1)^2 \right] \nonumber\\
&+ r(t)^3 \epsilon \cos^2\theta(t) \sin \theta(t) \left[ \sigma ^2\tau (\tau -1)^3 \right] \nonumber \\  &- r(t)^{2} \cos^2\theta(t) \left[ (1/2)(1 + 2a\sigma(\tau - 1)(\tau - 1)\tau)\right], \nonumber \qedhere
\label{eq29}
\end{align}
%\subsection{\label{Third}{The equation for phase evolution}}
We assume that $r(t)$ and angular velocity($\omega$) remain constant along the limit cycle. For examining these assumptions, we will try to find the expression for $\dot{\theta}(t)$ and $\dot{r}(t)$. By getting the time derivative of the equation~\ref{eq27}, we can find $\dot{r}$ and $\dot{\theta}$ as follows:
\begin{eqnarray}
&& \dot{\theta}(t) = \dfrac{-\dot{dx}}{r(t)\sin\theta(t)} + \dfrac{\dot{r}(t)}{r(t)}\dfrac{\cos\theta(t)}{\sin\theta(t)},
\label{eq30}
\end{eqnarray}
\begin{eqnarray}
&& \dot{r}(t) = \dfrac{\dot{dx}}{\cos\theta(t)} + r(t)\dot{\theta}(t)\dfrac{\sin\theta(t)}{\cos\theta(t)},
\label{eq31}
\end{eqnarray}
and
\begin{eqnarray}
&& \dot{\theta}(t) = \dfrac{-\dot{dy}}{r(t) \epsilon\cos\theta(t)} - \dfrac{\dot{r}(t)}{r(t)}\dfrac{\sin\theta(t)}{\cos\theta(t)},
\label{eq32}
\end{eqnarray}
\begin{eqnarray}
&& \dot{r}(t) = \dfrac{\dot{dy}}{\epsilon\sin\theta(t)}-r(t)\dot{\theta}(t) \dfrac{\cos\theta(t)}{\sin\theta(t)}.
\label{eq33}
\end{eqnarray}
By substituting equations~\ref{eq25} and~\ref{eq26} in equations~\ref{eq30} and~\ref{eq32} we obtain the following equations for phase evolution:
\begin{align}
\dot{\theta}(t) &= \epsilon - \dfrac{\cos \theta(t)}{\sin \theta(t)} c_{6}+ r(t)\epsilon \cos\theta(t)c_{5} \\
& + r^2(t)\epsilon \cos^2\theta(t)c_{4} - r(t)^{2} \dfrac{\cos^2\theta(t)}{\sin \theta(t)} c_{3} \nonumber \\
&- r(t)^{2} \dfrac{\cos^3\theta(t)}{\sin \theta(t)} c_{2}+ \dfrac{\dot{r}(t)}{r(t)}\dfrac{\cos\theta(t)}{\sin\theta(t)}, \nonumber \qedhere
\label{eq34}
\end{align}

\begin{align}
\dot{\theta}(t) &= \dfrac{1}{\epsilon}c_{7} + r(t)\sin\theta(t)c_{5}\\
&+ r^2(t)\cos\theta(t)\sin\theta(t)c_{4} - r(t)^{2} \dfrac{1}{\epsilon} \cos^2\theta(t)c_{2} \nonumber \\
&-r(t)\dfrac{1}{\epsilon}\cos\theta(t)c_{8}- \dfrac{\dot{r}(t)}{r(t)}\dfrac{\sin\theta(t)}{\cos\theta(t)}, \nonumber \qedhere
\label{eq35}
\end{align}
where the coefficients in both equations are as following:
\begin{eqnarray}
&& c_{1} = \sigma^{3} (\tau - 1)^{4}\tau^{2},  \\
&& c_{2} = (\sigma /2)(1 + 2a\sigma(\tau - 1))(\tau - 1)^{2}\tau^{2}, \nonumber \\
&& c_{3} = (1/2)(-1 + (1 + 2a\sigma(\tau - 1))(\tau - 1)\tau), \nonumber \\
&& c_{4} = \sigma^{2}(\tau - 1)^{3}\tau,\nonumber \\
&& c_{5} = \sigma(\tau - 1)^{2}, \nonumber \\
&& c_{6} = (1/\sigma)(1/2 + 1/(\tau - 1)) + a\tau, \nonumber \\
&& c_{7} = a-\dfrac{1}{2\sigma} -a\sigma, \nonumber \\
&& c_{8} = (1/2)(1 + 2a\sigma(\tau - 1)(\tau - 1)\tau). 
\nonumber
\label{eq36}
\end{eqnarray}
If we suppose that we have an ellipse with a constant angular velocity, then the period would be $T = \dfrac{2\pi}{\omega}$, where $\dot{\theta}(t) \simeq \omega$ can be obtained by calculating the following integral:
\begin{eqnarray}
&& \int_{0}^{2\pi} \dfrac{d\theta}{\dot{\theta}(t)} = \int_{0}^{T} dt = T
\label{eq37}
\end{eqnarray} 
By comparing equations~\ref{eq34} and~\ref{eq35}, and setting the constant parts of them equal, the following constrain is obtained:
\begin{eqnarray}
&& \epsilon = \dfrac{1}{\epsilon}(a-\dfrac{1}{2\sigma}- a\tau),
\label{eq38}
\end{eqnarray}
therefore:
\begin{eqnarray}
&& \dot{\theta} = \omega = \epsilon = \pm \sqrt{a(1-\tau)- \dfrac{1}{2\sigma}}.
\label{eq39}
\end{eqnarray}

Thus, we found the frequency along the limit cycle which is in agreement with the simulation results.

%\subsection{\label{Fourth}{The equation for radius evolution}}

To find the radius of the limit cycle, one can combine equations~\ref{eq30} and~\ref{eq33}, which leads to:
\begin{eqnarray}
&& \dot{r}(t)(\dfrac{1}{\cos(\epsilon t)}) = \dfrac{d\dot{x}}{\cos(\epsilon t)}+\dfrac{\sin(\epsilon t)d\dot{y}}{\epsilon \cos(\epsilon t)^2},
\label{eq40}
\end{eqnarray}
In the next step, we substitute $\dot{dx}$ and $\dot{dy}$ values from equations~\ref{eq28} and~\ref{eq29}, which results to:
\begin{align}
\dot{r}(t) &= \dfrac{1}{2\sigma (-1+(\sigma (\tau-1) \tau)r(t) \cos(\epsilon t))}\left[ \\
& r(t) \cos(\epsilon t)^2(-(1 +\tau)/(\tau-1)-2 a\sigma \tau) \nonumber \\
& +r(t)^2 \cos(\epsilon t)^3 (\sigma + 2 \sigma \tau + 2a\sigma^2(\tau-1)\tau) \nonumber \\
& + r(t)^3 \cos(\epsilon t)^4 (-\sigma^2(\tau-1)\tau) \nonumber \\
& + r(t) \cos(\epsilon t)\sin(\epsilon t)((2\epsilon^2 \sigma +1+ 2a\sigma(\tau - 1))/\epsilon)\nonumber \\
& + r(t)^2 \cos(\epsilon t)^2 \sin(\epsilon t)(-2 \epsilon \sigma^2(\tau-1)) \nonumber \\
& + r(t)^2 \cos(\epsilon t) \sin(\epsilon t)^2(-2\sigma^2(\tau- 1)^2)\right] ,\nonumber \qedhere 
\label{eq41}
\end{align}
We can see that $\dot{r}(t)$ is oscillating and equals to zero at $t = k\pi/2\epsilon$. Therefore, the constraint $\dot{r}(t = 2k\pi/\epsilon) =\dot{r}(t = (2k+1)\pi/\epsilon)  $ holds, which results to the following approximate value for radius:
\begin{eqnarray}
&& r = \dfrac{2a\sqrt{\dfrac{\gamma \tau(1+\tau)}{1+\gamma-\tau+\gamma \tau}}}{1+\gamma+\tau+\gamma \tau}
\end{eqnarray}

\subsection{\label{TWO}{Demographic noise strength derivation}}
%The attempt to find an analytic solution for the stochastic RMA model, in the stationary state, is explained here in the same order of the steps we took. These results are an analytic approximation, of the stochastic model. %In subsection~\ref{Fifth}
First, the general method for finding the stationary probability distribution and the Langevin equation of a typical non-linear system is described. Then, %in subsection~\ref{Sixth}
the same method is applied on RMA model. Next, 
%in subsection~\ref{Seventh}
we found both deterministic and stochastic terms of the Fokker-Planck equation of the system, by writing a master equation and expanding it around fixed point. Finally, 
%in subsection~\ref{Eighth}, 
we tried to write the equivalent stochastic term on the stable limit cycle. We found expressions for noise terms in polar coordinates depending on the noise terms in Cartesian coordinates, which are found by expanding master equation around fixed point. With this method, the approximate noise strength will be found. 
%\subsection{\label{Fifth}{The Stationary Probability and Langevin Equation of the Non-Linear Systems in General}}

As a general method for finding the stationary distribution of a given non-linear system, one needs to write the Fokker-Planck equation (FPE) corresponding to the system, and apply the Van-Kampen system size expansion around the deterministic solution. 
If we have a set of non-linear equations, as below:
\begin{eqnarray} 
%\begin{split}
&& \dot{x_{1}} = F_{1}(x_{1},x_{2}),  \\ 
&& \dot{x_{2}} = F_{2}(x_{1},x_{2}), \nonumber
%\end{split}
\label{eq01}
\end{eqnarray}
In a stochastic regime, where noise is added, equations change as following:
\begin{eqnarray} 
%\begin{split}
&& \dot{x_{1}} = F_{1}(x_{1},x_{2})+\eta_{1}(t),  \\ 
&& \dot{x_{2}} = F_{2}(x_{1},x_{2})+\eta_{2}(t), \nonumber
%\end{split}
\label{eq02}
\end{eqnarray}
Where $\eta(t)$ is a Gaussian white noise, which has the following properties:
\begin{eqnarray}
&& \left\langle\eta(t)\right\rangle  = 0 \\
&& \left\langle\eta(t_{1}) \eta(t_{2})\right\rangle  = \beta^{-1} \delta(t_{1}-t_{2}) \nonumber
\label{eq03}
\end{eqnarray}
The evolution of the $p(x,t)$ will have the Fokker-Planck 
equation in form of:
\begin{eqnarray}
\partial_{t}P(x,t) = \left( -\sum_{i} \partial_{i}F_{i}(x)+\dfrac{1}{2}\beta^{-1}\sum_{i}\partial_{i}^{2}\right)  P(x.t),
\label{eq04}
\end{eqnarray}
When the noise amplitude depends on the system parameters, the FPE takes the following form:
\begin{eqnarray}
\partial_{t}P(x,t) = \left( -\sum_{i} \partial_{i}F_{i}(x)+\dfrac{1}{2}\sum_{i}\partial_{i}^{2}B_{ij}\right)  P(x.t),
\label{eq04}
\end{eqnarray}
In the weak noise limit, Van Kampen expansion in terms of the square root of the noise amplitude can be applied, such that:
\begin{eqnarray}
&& y(t) = \beta^{1/2}\left[ x-x_{det}(t)\right] \\
&& F(x) = F(x_{det})+(\dfrac{\partial F}{\partial x})_{x_{det}}.\beta^{-1/2}y, \nonumber
\label{eq05}
\end{eqnarray}
then FPE takes the following form~\cite{kurrer1991effect}:
\begin{eqnarray}
\partial_{t}P(y,t)= -\sum_{i,j}A_{ij}(t)\dfrac{\partial}{\partial y_{i}}y_{i}P(y,t)+\dfrac{1}{2}\sum_{i,j}\tilde{B}_{ij}\dfrac{\partial^{2}}{\partial y_{i} \partial y_{j}}P(y,t),
\label{eq06}
\end{eqnarray}
where:
\begin{eqnarray}
&& A_{ij} = \dfrac{\partial F_{i}}{\partial x_{j}}\vert_{\vec{x}_{det}},  \\
&& \tilde{B}_{ij} = B_{ij}\vert_{\vec{x}_{det}}. \nonumber  
\label{eq20}
\end{eqnarray}
%Then the solution for $P(y,t)$ would be a Gaussian distribution as following:
%\begin{eqnarray}
%P(y,t) = \dfrac{1}{2\pi \sqrt{Det C}} exp (-\dfrac{1}{2}y^{T}.C^{-1}.y),
%\label{eq08}
%\end{eqnarray}
%where: 
%\begin{eqnarray}
%&& C_{ij} = \langle y_{i}(t)y_{j}(t)\rangle,\\ 
%&& \sigma = \sqrt{Tr(C)}, \nonumber
%\label{eq09}
%\end{eqnarray}
%$C(t)$ is determined by the following equation of motion:
%\begin{eqnarray}
%\dot{C} = A(t)C + CA^{T} + B.
%\label{eq10}
%\end{eqnarray}

%So by knowing the $A$ and $B$ with their elements, we can find the $C$ and then, we will have a complete description of the probability distribution[1].
The corresponding Langevin equation for $y_{i}$ can be obtained as following:
\begin{eqnarray}
&& \dot{y_{i}} = \sum_{j} A_{i,j} + \eta_{i},
\label{eq50}
\end{eqnarray} 
where the equation $\left\langle \eta_{i}^{2} \right\rangle = B_{ii}  $ holds.
%\subsection{\label{Sixth}{Finding A and B by Expanding the Master Equation}}

In order to apply the mentioned method on our system, we start from the RMA equations for population dynamics:
\begin{eqnarray}
&& \dot{R} = aR-\dfrac{R^2}{2N}- \dfrac{sRF}{1+ s \tau R},  \\  
&& \dot{F} = -F+\dfrac{sRF}{1+ s \tau R},\nonumber
\label{eq13}
\end{eqnarray} 

%Scaled by System Size: ($x = R/N, y = F/N$)
%\begin{eqnarray} 
%%\begin{split}
%&& \dot{x} = ax - \dfrac{x^2}{2}- \dfrac{\sigma xy}{1+ \sigma \tau x},  \\ 
%&& \dot{y} = -y + \dfrac{\sigma xy}{1+ \sigma \tau x}, \nonumber
%%\end{split}
%\label{eq14}
%\end{eqnarray}
   
One can obtain the stochastic master equation, in which in addition to the rates of death and birth and predation, there is a probability of no change in population numbers in a given time step in the systemm which can be obtained as following:
\begin{align}
\dfrac{P(R,F,t+\delta t)-P(R,F,t)}{\delta t} &= a(R-1)P(R-1,F,t)\\
&+\dfrac{1}{2N}(R+1)^2P(R+1,F,t)  \nonumber \\
& +\dfrac{s(R+1)(F-1)}{1+s\tau R}P(R+1,F-1,t)\nonumber \\
&+(F+1)P(R,F+1,t) \nonumber \\
&-(aR+\dfrac{R^2}{2N}+\dfrac{sRF}{1+s\tau R}+F)P(R,F,t), \nonumber 
\label{eq15}
\end{eqnarray}

To investigate the system in continuous limit, we scaled the master equation by the system size($x = R/N, y = F/N$):

\begin{align}
\dfrac{P(x,y,t+\delta t)-P(x,y,t)}{\delta t} &= Na(x-\dfrac{1}{N})P(x-\dfrac{1}{N},y,t)\\
& +\dfrac{N}{2}(x+\dfrac{1}{N})^2P(x+\dfrac{1}{N},y,t) \nonumber  \\
& +N\dfrac{\sigma (x+\dfrac{1}{N})(y-\dfrac{1}{N})}{1+\sigma \tau x}P(x+\dfrac{1}{N},y-\dfrac{1}{N},t) \nonumber \\
&+N(y+\dfrac{1}{N})P(x,y+\dfrac{1}{N},t)\nonumber \\
& -N(ax+\dfrac{x^2}{2}+\dfrac{\sigma xy}{1+\sigma \tau x}+y)P(x,y,t). \nonumber
\label{eq16}
\end{align}

Next, we applied the Taylor expansion on the continuous master equation around $(x,y)$ (where later is going to be substituted by the values of the fixed point):
\begin{align}
\dfrac{\partial P(x,y,t)}{\partial t}&= N\left[ axP(x,y)-\dfrac{1}{N} \dfrac{\partial }{\partial x} axP(x,y)+\dfrac{1}{2N^2}\dfrac{\partial^2}{\partial x^2} axP(x,y)\right]   \\ 
&+N\left[ \dfrac{x^2}{2}P(x,y)+\dfrac{1}{N}\dfrac{\partial}{\partial x}\dfrac{x^2}{2}P(x,y)+\dfrac{1}{2N^2}\dfrac{\partial^2}{\partial x^2}\dfrac{x^2}{2}P(x,y)\right]  \nonumber \\ 
& + N\left[ \dfrac{\sigma xy}{1+\sigma \tau x}P(x,y)+ \dfrac{1}{N}\dfrac{\partial}{\partial x}\dfrac{\sigma xy}{1+\sigma \tau x}P(x,y) - \dfrac{1}{N}\dfrac{\partial}{\partial y}\dfrac{\sigma xy}{1+\sigma \tau x}P(x,y)\right] \nonumber \\
& + N\left[ \dfrac{1}{2N^2}\dfrac{\partial^2}{\partial x^2}\dfrac{\sigma xy}{1+\sigma \tau x}P(x,y)+\dfrac{1}{2N^2}\dfrac{\partial^2}{\partial y^2}\dfrac{\sigma xy}{1+\sigma \tau x}P(x,y)\right]  \nonumber \\
& + N\left[ - \dfrac{1}{2N^2}\dfrac{\partial^2}{\partial x\partial y}\dfrac{\sigma xy}{1+\sigma \tau x}P(x,y) -\dfrac{1}{2N^2}\dfrac{\partial^2}{\partial y \partial x}\dfrac{\sigma xy}{1+\sigma \tau x}P(x,y)\right]  \nonumber \\ 
&+ N\left[  yP(x,y) +  \dfrac{1}{N}\dfrac{\partial}{\partial y} yP(x,y) + \dfrac{1}{2N^2}\dfrac{\partial^2}{\partial y^2} yP(x,y)\right] \nonumber \\ 
&- N\left[(ax + \dfrac{x^2}{2} + \dfrac{\sigma xy}{1+\sigma \tau x} + y) P(x,y)\right] , \nonumber
\label{eq17}
\end{align}
Therefore, the classical form of the Fokker-Planck equation can be found as following:
\begin{align}
\partial_{t}P(x,y) &= \dfrac{\partial}{\partial x} (-ax + \dfrac{x^{2}}{2} + \dfrac{\sigma xy}{1+\sigma \tau x})P(x,y)\\
&  + \dfrac{\partial}{\partial y}(-\dfrac{\sigma xy}{1+\sigma \tau x} + y )P(x,y)\nonumber \\
& + \dfrac{1}{2N}\dfrac{\partial^{2}}{\partial x^{2}}(ax + \dfrac{x^{2}}{2} + \dfrac{\sigma xy}{1+\sigma \tau x})P(x,y) \nonumber \\
&+ \dfrac{1}{2N}\dfrac{\partial^{2}}{\partial y^{2}}(\dfrac{\sigma xy}{1+\sigma \tau x} + y)P(x,y) \nonumber \\
& -\dfrac{1}{2N}\dfrac{\partial^{2}}{\partial x \partial y}(\dfrac{\sigma xy}{1+\sigma \tau x})P(x,y) \nonumber \\
&- \dfrac{1}{2N}\dfrac{\partial^{2}}{\partial y \partial x}(\dfrac{\sigma xy}{1+\sigma \tau x})P(x,y), \nonumber
\label{eq18}
\end{align}
Where the coefficients are defined as follows:
\begin{eqnarray}
&& F_{x} = ax-\dfrac{x^{2}}{2}-\dfrac{\sigma xy}{1+\sigma \tau x},  \\
&& F_{y} = \dfrac{\sigma xy}{1+\sigma \tau x} -y,\nonumber \\
&& B_{xx} = \dfrac{1}{N}(ax + \dfrac{x^{2}}{2} + \dfrac{\sigma xy}{1+\sigma \tau x}), \nonumber \\
&& B_{yy} = \dfrac{1}{N}(\dfrac{\sigma xy}{1+\sigma \tau x} + y), \nonumber \\
&& B_{xy} = - \dfrac{1}{N}(\dfrac{\sigma xy}{1+\sigma \tau x}), \nonumber \\
&& B_{yx} = - \dfrac{1}{N}(\dfrac{\sigma xy}{1+\sigma \tau x}). \nonumber
\label{eq360}
\end{eqnarray}

As discussed before, the FPE after the Van-Kampen expansion will take the bellowing form:
\begin{eqnarray}
\partial_{t}P(x,t)= -\sum_{i,j}A_{ij}(t)\dfrac{\partial}{\partial x_{i}}x_{i}P(x,t)+\dfrac{1}{2}\sum_{i,j}\tilde{B}_{ij}\dfrac{\partial^{2}}{\partial x_{i} \partial x_{j}}P(y,t),
\label{eq19}
\end{eqnarray}
where: %the corresponding terms can be obtained by knowing equivalent terms to the FPE before expansion using the following equations:
\begin{eqnarray}
&& A_{ij} = \dfrac{\partial F_{i}}{\partial x_{j}}\vert_{(x,y)_{det}},  \\
&& \tilde{B}_{ij} = B_{ij}\vert_{(x,y)_{det}} \nonumber  
\label{eq20}
\end{eqnarray}

%knowing the values for the matrix B from expanding the Van-Kampen expansion of the Fokker-Planck equations:
%\begin{eqnarray}
%&& B_{xx} = \dfrac{1}{N}\left( ax + \dfrac{x^{2}}{2} + \dfrac{xy\sigma}%{1+x\tau\sigma}\right),  \\
%&& B_{yy} = \dfrac{1}{N}\left( y + \dfrac{xy\sigma}{1+x\tau\sigma}%\right), \nonumber \\
%&& B_{xy} = B_{yx} = -\dfrac{1}{N}\left( \dfrac{xy\sigma}{1+x\tau\sigma}\right). \nonumber
%\label{eq55}
%\end{eqnarray}
%So by substituting the values for fixed point(below), we have the $A$ %and $B$ matrix depending on the system parameters.
%\begin{equation}
%(x_{0}, y_{0}) \rightarrow (\frac{1}{\sigma (1-\tau)}, \dfrac{2a\sigma %(1-\tau)-1 }{2\sigma^2 (1-\tau)^2})
%\label{eq21}
%\end{equation}

%\subsection{\label{Seventh}{Determining the Noise term on Limit Cycle Regime}}

As we mentioned before, we suppose that our deterministic solution is an ellipse. Then add noise terms on the horizontal and normal directions of the limit cycle, as follows:
\begin{eqnarray}
&& \vec{y} = \vec{y}_{det} + \begin{bmatrix}
r(t)\cos\theta(t)\\
\epsilon r(t) \sin\theta(t)
\end{bmatrix},
\label{eq45}
\end{eqnarray}
which can  be transformed to the following equation:
\begin{eqnarray}
&& \vec{y} = \vec{y}_{det} + (r^{*}(t)+ \dfrac{\xi_{r}}{\sqrt{N}})\begin{bmatrix}
\cos(\omega t +\dfrac{\xi_{\theta}}{\sqrt{N}})\\
\epsilon \sin(\omega t +\dfrac{\xi_{\theta}}{\sqrt{N}})
\end{bmatrix},
\label{eq46}
\end{eqnarray}
while $r^{*}$ and $\omega$ are constants of the stable limit cycle. By expanding around $\omega t$ we find that:
\begin{eqnarray}
&& \vec{y} = \vec{y}_{det}+ \begin{bmatrix}
r^{*}(t)\cos(\omega t)\\
\epsilon r^{*}(t)\sin(\omega t)
\end{bmatrix} +(\dfrac{1}{\sqrt{N}}) \begin{bmatrix}
\xi_{r}\cos(\omega t)-\xi_{\theta} r^{*}(t)\sin(\omega t)\\
\xi_{r}\epsilon \sin(\omega t) + \xi_{\theta}\epsilon r^{*}\cos(\omega t)
\end{bmatrix},
\label{eq47}
\end{eqnarray}
where:
\begin{eqnarray}
&& \xi_{x} = \xi_{r}\cos(\omega t) - \xi_{\theta}r^{*}(t)\sin(\omega t), \\
&& \xi_{y} = \xi_{r}\epsilon \sin(\omega t) - \xi_{\theta}\epsilon r^{*}(t)\cos(\omega t), \nonumber 
\label{eq48}
\end{eqnarray}
The Langvin equation corresponding to the $\xi_{x}$ and $\xi_{y}$ is written as follows:
\begin{eqnarray}
&& \dot{\xi_{x}} = \sum_{i} A_{x,i} + \eta_{x}  \\
&& \dot{\xi_{y}} = \sum_{i} A_{y,i} + \eta_{y} \nonumber
\label{eq50}
\end{eqnarray} 
we can straightforwardly express $\xi_{n}$ and $\xi_{l}$ as follows:
\begin{eqnarray}
&& \xi_{r} = \cos(\omega t) \xi_{x} + \dfrac{1}{\epsilon} \sin(\omega t)\xi_{y},  \\
&& \xi_{\theta} = \dfrac{-1}{r^{*}(t)}\sin(\omega t) \xi_{x} + \dfrac{1}{r^{*}(t)\epsilon}\cos(\omega t)\xi_{y}. \nonumber
\label{eq49}
\end{eqnarray}
%\subsection{\label{Eighth}{Substituting Known x-y Noise Terms}}
Getting the time derivation of the above equations leads to:
\begin{eqnarray}
&& \dot{\xi_{r}}=  -\epsilon \sin(\epsilon t) \xi_{x} +  \cos(\epsilon t)\dot{\xi_{x}}+ \cos(\epsilon t) \xi_{y} +\dfrac{1}{\epsilon}\sin(\epsilon t)\dot{\xi_{y}},  \\
&& \dot{\xi_{\theta}}=  -\dfrac{1}{r^{*}} \epsilon \cos(\epsilon t) \xi_{x} -\dfrac{1}{r^{*}}\sin(\epsilon t)\dot{\xi_{x}}- \dfrac{1}{r^{*}} \sin(\epsilon t) \xi_{y}  +\dfrac{1}{r^{*}\epsilon}\cos(\epsilon t)\dot{\xi_{y}}. \nonumber
\label{eq51}
\end{eqnarray}
Expecting from phase evolution equation for Brownian motion, one knows that:
\begin{align}
    & \dot{\theta} = [...]_{det} + \sqrt{Dt}, \\
    & \dot{\theta}^{2} = [...]_{det}^{2} + 2[...]_{det}\sqrt{Dt} + Dt. \nonumber
\end{align}
Therefore, the noise strength can be obtained from the coefficient of $t$ in the equation for second power of $\xi_{\theta}$.  
Combination of the equations~\ref{eq50} and~\ref{eq51} and separating the terms that are purely stochastic gives rise to the:
\begin{align}
\dot{\xi}_{\theta}^{2}  &= [...]_{det+stoch} + \dfrac{1}{r^{*}^{2}} \sin^2(\epsilon t) \eta_{x}^{2}  \\
&+\dfrac{1}{r^{*}^{2} \epsilon^2} \cos^2(\epsilon t) \eta_{y}^{2} - \dfrac{2}{r^{*}\epsilon^2} \sin(\epsilon t)\cos(\epsilon t) \eta_{x} \eta_{y},  \nonumber
\label{eq52}
\end{align}
where the term $[...]_{det+stoch}$ summarizes the combination of the stochastic and deterministic terms as in $[...]_{det}^{2} + 2[...]_{det}\sqrt{Dt} $. Now if we take an average over one period of time, the stochastic term will change as following:

\begin{align}
\left\langle \dot{\xi}_{\theta}^{2} \right\rangle  &= [...]_{det+stoch} + \dfrac{1}{r^{*}^{2}}\left\langle \sin^2(\epsilon t)\right\rangle \left\langle \eta_{x}^{2} \right\rangle \\
&+\dfrac{1}{r^{*}^{2} \epsilon^2}\left\langle \cos^2(\epsilon t)\right\rangle \left\langle \eta_{y}^{2}\right\rangle - \dfrac{2}{r^{*}\epsilon^2} \left\langle \sin(\epsilon t)\cos(\epsilon t)\right\rangle  \left\langle \eta_{x} \eta_{y}\right\rangle,  \nonumber \\
\left\langle \dot{\xi}_{\theta}^{2} \right\rangle  &= [...]_{det+stoch} + \dfrac{1}{r^{*}^{2}}\dfrac{t}{2} \left\langle \eta_{x}^{2} \right\rangle +\dfrac{1}{r^{*}^{2} \epsilon^2}\dfrac{t}{2}\left\langle \eta_{y}^{2}\right\rangle. \nonumber
\label{eq53}
\end{align}
From previous calculations we had:
\begin{eqnarray}
&& \left\langle \eta_{x}^{2} \right\rangle = B_{xx},  \\
&& \left\langle \eta_{y}^{2}\right\rangle = B_{yy}, \nonumber \\
&& \left\langle \eta_{x} \eta_{y}\right\rangle = B_{xy}. \nonumber
\label{eq54}
\end{eqnarray}
Substituting the values for $\eta^{2}$ in equation~\ref{eq53}, one finds that:
\begin{align}
 \left\langle \xi_{\theta}^{2} \right\rangle &= [...]_{det+stoch} + \dfrac{1}{r^{*}^{2}}(\dfrac{t}{2})\left[ \dfrac{1}{N}\left(ax + \dfrac{x^{2}}{2} + \dfrac{xy\sigma}{1+x\tau\sigma}\right)\right]\\
 &+\dfrac{1}{r^{*}^{2}\epsilon^{2}}(\dfrac{t}{2})\left[ \dfrac{1}{N}\left( y + \dfrac{xy\sigma}{1+x\tau\sigma}\right)\right]\nonumber 
\label{eq55}
\end{align}
As in $\dot{\xi_{\theta}}^{2} = [...]_{det+stoch} + Dt$, we look for the coefficient of the $t$, as a result the expression for D, depends on $x$ and $y$ which is expected, as it supposed to be a demographic noise. By substituting the value for parameter $r^{*}$ from previous subsection, one can find the approximate value for $D$ as folllows:
\begin{eqnarray}
&& D \sim \dfrac{1}{2r^{*}^{2}\epsilon^{2}N}\left[\epsilon^{2} ( ax + \dfrac{x^{2}}{2} + \dfrac{xy\sigma}{1+x\tau\sigma})+ y+\dfrac{xy\sigma}{1+x\tau\sigma}\right]. 
\label{eq58}
\end{eqnarray}